\pdfoutput=1

\documentclass[11pt,twoside,a4paper,cmspaper,final,collab]{cms-tdr}
\def\svnVersion{d0e6c88-D}\def\svnDate{2020/05/05}\def\cmsCernNoTag{CERN-EP-2019-233}\def\cmsCernDate{\today}\def\cmsMessage{"Published in the Journal of High Energy Physics as \href{http://dx.doi.org/10.1007/JHEP05(2020)032}{\doi{10.1007/JHEP05(2020)032}.}"}

\begin{document}\cmsNoteHeader{SUS-19-009}

\hyphenation{had-ron-i-za-tion}
\hyphenation{cal-or-i-me-ter}
\hyphenation{de-vices}
\newcommand{\pp}{\ensuremath{\Pp{}\Pp{}}\xspace}
\newcommand{\tmod}{\ensuremath{t_{\text{mod}}}\xspace}
\newcommand{\Mlb}{\ensuremath{M_{\ell{\PQb}}}\xspace}
\newcommand{\MT}{\ensuremath{M_{\mathrm{T}}}\xspace}
\newcommand{\NJ}{\ensuremath{N_{\mathrm{j}}}\xspace}
\newcommand{\NISR}{\ensuremath{N_{\mathrm{j}}^\mathrm{ISR/FSR}}\xspace}
\newcommand{\NB}{\ensuremath{N_{{\PQb}}}\xspace}
\newcommand{\Nsoftb}{\ensuremath{N_{\PQb,\text{ soft}}}\xspace}
\newcommand{\Nmedb}{\ensuremath{N_{\PQb,\text{ med}}}\xspace}
\newcommand{\ptsum}{\ensuremath{\pt^{\text{sum}}}\xspace}
\newcommand{\minDPhiMETjet}{\ensuremath{\min\Delta\phi(j_{1,2},\ptvecmiss)}\xspace}
\newcommand{\wjets}{\ensuremath{\PW+\text{jets}}\xspace}
\newcommand{\zjets}{\ensuremath{\PZ+\text{jets}}\xspace}
\newcommand{\gjets}{\ensuremath{\Pgg+\text{jets}}\xspace}
\newcommand{\ttjets}{\ensuremath{\ttbar+\text{jets}}\xspace}
\newcommand{\Znunu}{\ensuremath{\PZ\to\nu\bar{\nu}}\xspace}
\newcommand{\Lint}{\ensuremath{137\fbinv}\xspace}
\newcommand{\lsp}{\PSGczDo}
\newcommand{\chgo}{\PSGcpmDo}
\providecommand{\NA}{\text{---}\xspace}
\newcommand{\cmsTable}[1]{\resizebox{\textwidth}{!}{#1}}
\newlength\cmsTabSkip\setlength{\cmsTabSkip}{1ex}
\providecommand{\CL}{CL\xspace}

\cmsNoteHeader{SUS-19-009}
\title{Search for direct top squark pair production in events with one lepton, jets, and missing transverse momentum at 13\TeV with the CMS experiment}

\date{\today}

\abstract{
A search for direct top squark pair production is presented. The search is based on proton-proton collision data at a center-of-mass energy of 13\TeV recorded by the CMS experiment at the LHC during 2016, 2017, and 2018, corresponding to an integrated luminosity of \Lint. The search is carried out using events with a single isolated electron or muon, multiple jets, and large transverse momentum imbalance. The observed data are consistent with the expectations from standard model processes. Exclusions are set in the context of simplified top squark pair production models. Depending on the model, exclusion limits at 95\% confidence level for top squark masses up to 1.2\TeV are set for a massless lightest supersymmetric particle, assumed to be the neutralino. For models with top squark masses of 1\TeV, neutralino masses up to 600\GeV are excluded.
}

\hypersetup{
pdfauthor={CMS Collaboration},
pdftitle={Search for direct top squark pair production in events with one lepton, jets and missing transverse momentum at 13 TeV},
pdfsubject={CMS},
pdfkeywords={CMS, supersymmetry, top squark}}

\maketitle

\section{Introduction}
\label{sec:intro}

Supersymmetry (SUSY)~\cite{Ramond:1971gb,Golfand:1971iw,Neveu:1971rx,Volkov:1972jx,Wess:1973kz,Wess:1974tw,Fayet:1974pd,Nilles:1983ge} is an attractive extension of the standard model (SM), characterized by the presence of SUSY partners for every SM particle.
These partner particles have the same quantum numbers as their SM counterparts, except for the spin, which differs by one-half unit.
In models with $R$-parity conservation~\cite{Farrar:1978xj}, the lightest supersymmetric particle (LSP) is stable, and, if neutral, could be a dark matter candidate~\cite{Jungman:1995df}.
The extended particle spectrum in SUSY scenarios allows for the cancellation of quadratic divergences arising from quantum corrections to the Higgs boson mass~\cite{tHooft:1979rat,Dine:1981za,Dimopoulos:1981au,Dimopoulos:1981zb,Kaul:1981hi}.
Scenarios realizing this cancellation often contain top squarks ($\PSQt$), SUSY partners of the SM top quark ($\PQt$), and higgsinos, SUSY partners of the SM Higgs boson, with masses near the electroweak scale. 
The {\PSQt} pair production cross section is expected to be large compared to the electroweak production of higgsinos at CERN LHC for $\PSQt$ masses near the electroweak scale.

\begin{figure*}[hbt]
\centering
  \includegraphics[width=0.32\textwidth]{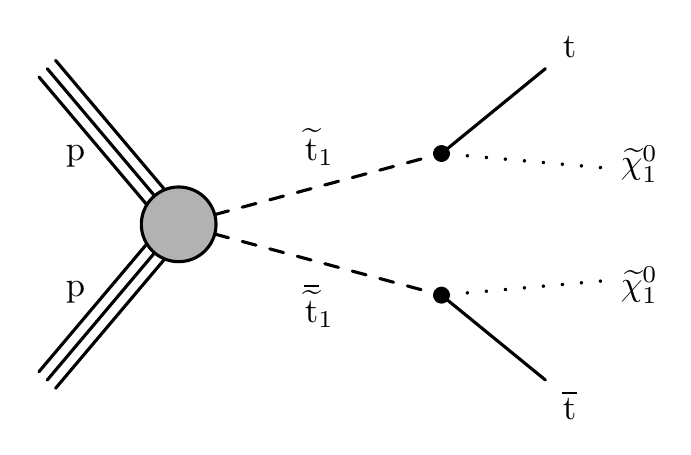}
  \includegraphics[width=0.32\textwidth]{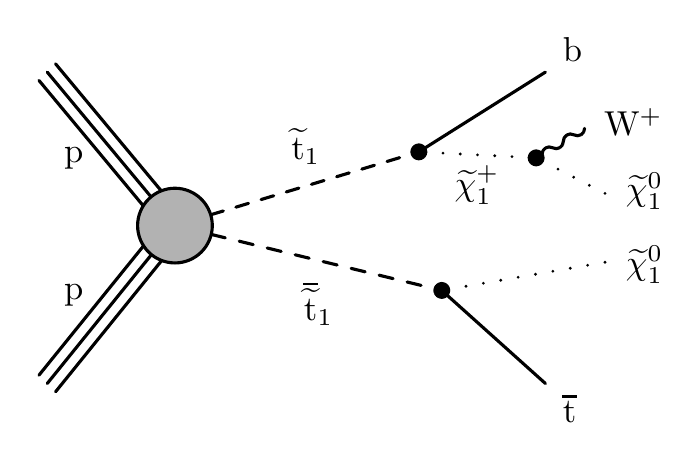}
  \includegraphics[width=0.32\textwidth]{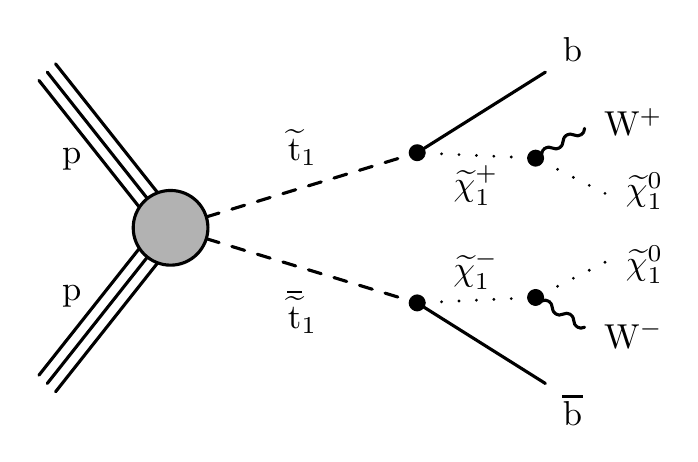}
  \caption{
        Diagrams for top squark pair production, with each \PSQt decaying either to $\PQt\lsp$ or to $\PQb\chgo$. For the latter decay, the \chgo decays further into a $\PW$ boson and a \PSGczDo.
  }
    \label{fig:diagram}
\end{figure*}

In this paper, a search is presented for top squark pair production in final states with events from $\pp$ collisions at $\sqrt{s}=13\TeV$, collected between 2016 and 2018 by the CMS experiment, corresponding to an integrated luminosity of \Lint.
Two top squark decay modes are considered: the decay to a top quark and the lightest neutralino ($\lsp$), which is taken to be the LSP, or the decay to a bottom quark ({\PQb}) and the lightest chargino ($\chgo$). In the latter scenario, it is assumed that the $\chgo$ decays to a $\PW$ boson and the $\lsp$.
The mass of the chargino is chosen to be $(m_{\PSQt} + m_{\lsp})/2$.
The corresponding diagrams are given in Fig.~\ref{fig:diagram}.
The common experimental signature for pair production with these decay modes is $\PW \PW ^{(*)} + \PQb \PQb + \lsp \lsp $.
The analysis is based on events where one of the \PW bosons decays leptonically and the other hadronically.
This results in the event selection of one isolated lepton, at least 2 jets, and large missing transverse momentum ($\ptmiss$) from the two neutralinos and the neutrino.

Dedicated searches for top squark pair production in 13\TeV proton-proton ($\pp$) collision events have been carried out by both the ATLAS~\cite{Aaboud:2016lwz,Aaboud:2016uth,Aaboud:2017ejf,Aaboud:2017nfd,Aaboud:2017ayj,Aaboud:2017opj,Aaboud:2017nmi,Aaboud:2017aeu,Aaboud:2018kya,Aaboud:2018zjf} and CMS~\cite{Khachatryan:2016pxa,Sirunyan:2016jpr,Khachatryan:2017rhw,Sirunyan:2017xse,Sirunyan:2017wif,Sirunyan:2017kiw,Sirunyan:2017pjw,Sirunyan:2017leh,Sirunyan:2018iwl,Sirunyan:2018omt,Sirunyan:2018rlj,Sirunyan:2018ell,Sirunyan:2019zyu} Collaborations.
The search presented here improves the previous one~\cite{Sirunyan:2017xse}
by adding the data collected in 2017 and 2018, resulting in approximately a factor of four increase in the size of the data sample.
In addition, new search regions have been added, which are sensitive to scenarios where the mass of the top squark is close to the sum of the masses of either the \lsp and the top quark, or the \lsp and the $\PW$ boson. These scenarios are referred to as compressed mass scenarios hereafter.
In addition, a method has been implemented to identify top quarks that decay hadronically, and also the background estimation techniques have been improved.
The paper
is organized as follows:
Section~\ref{sec:CMS} and \ref{sec:mc} describe the CMS detector and the simulated samples used in this analysis. The object reconstruction and search strategy are presented in Section~\ref{sec:evtsel}.
The background prediction methods are described in Section~\ref{Sec:BkgEst}, and the relevant systematic uncertainties are discussed in Section~\ref{sec:syst}.
Results and interpretations are detailed in Section~\ref{sec:results}, and a summary is presented in Section~\ref{sec:summary}.

\section{The CMS detector}
\label{sec:CMS}
The central feature of the CMS apparatus is a superconducting solenoid of 6\unit{m} internal diameter, providing a magnetic field of 3.8\unit{T}. Within the solenoid volume are a silicon pixel and strip tracker, a lead tungstate crystal electromagnetic calorimeter (ECAL), and a brass and scintillator hadron calorimeter (HCAL), each composed of a barrel and two endcap sections. Forward calorimeters extend the pseudorapidity ($\eta$) coverage provided by the barrel and endcap detectors.
Muons are detected in gas-ionization chambers embedded in the steel flux-return yoke outside the solenoid.

Events of interest are selected using a two-tier trigger system.
The first level, composed of custom hardware processors, uses information from the calorimeters and muon detectors to select events in a fixed time interval of less than 4\mus.
The second level, called the high-level trigger, further decreases the event rate from around 100\unit{kHz} to less than 1\unit{kHz} before data storage.
A more detailed description of the CMS detector, together with a definition of the coordinate system used and the relevant kinematic variables, can be found in Refs.~\cite{Chatrchyan:2008zzk,Khachatryan:2016bia}.
The pixel tracker was upgraded before the start of the data taking period in 2017,
providing one additional layer of measurements compared to the older tracker~\cite{phase1trackerTDR}.

\section{Simulated samples}
\label{sec:mc}
{\tolerance=1200
Monte Carlo (MC) simulation is used to design the search, to aid in the estimation of SM backgrounds, and to evaluate the sensitivity of the analysis to top squark pair production.
Samples of events of SM $\ttbar$, $\wjets$, $\zjets$, and $\gjets$ processes and simplified SUSY top squark pair
production models are generated at leading-order (LO) in quantum chromodynamics (QCD) using the \MGvATNLO~2 (2.2.2 or 2.4.2) generator~\cite{Alwall:2014hca}.
The \MGvATNLO at next-to-LO (NLO) in QCD is used to generate samples of $\ttbar\PZ$, $\PW\PZ$, and $\ttbar\PW$ events, while single top quark events are generated at NLO in QCD using the \POWHEG~2.0~\cite{Nason:2004rx,Frixione:2007vw,Alioli:2010xd,Re:2010bp} program. Samples
of $\wjets$, \ttbar, and SUSY events are generated with four, three, and two additional partons included
in the matrix element calculations, respectively.
\par}

Since the data
used for this search were collected in three distinct periods (2016, 2017, and 2018), different
detector MC simulations are used to reflect the
running conditions. In addition, in some cases, the
generator settings are also different as described
below.

The NNPDF3.0~\cite{Ball:2011uy,Ball:2014uwa} parton distribution functions (PDFs) are used to generate all 2016 MC samples, while NNPDF3.1~\cite{Ball:2017nwa} is used for 2017 and 2018 samples.
The parton shower and hadronization are modeled with \PYTHIA~8.2 (8.205 or 8.230)~\cite{Sjostrand:2014zea}.  The MLM~\cite{Alwall:2007fs} and FxFx~\cite{Frederix:2012ps} prescriptions are employed to match
partons from the matrix element calculation to those from the parton showers, for the LO and NLO samples, respectively.

The 2016 MC samples are generated with the
\textsc{CUETP8M1}~\cite{Khachatryan:2015pea} \PYTHIA\ tune.
For the later
running periods, the \textsc{CP5}~\cite{Sirunyan:2019dfx} tune was used
for SM samples, and the SUSY samples use LO PDFs, combined with tune \textsc{CP2}, in order to avoid large negative weights that arise from PDF interpolations at very large energies.
The differences in jet kinematic properties between the SUSY and SM samples are due to different \PYTHIA\ tunes and are within 5\% of each other.
The \GEANTfour~\cite{Agostinelli2003250} package is used to simulate the response of the CMS detector for all SM processes, while the CMS fast simulation program~\cite{Abdullin:2011zz,Giammanco:2014bza} is used for SUSY samples.

Cross section calculations performed at next-to-NLO (NNLO) in QCD are used to normalize
the MC samples of $\wjets$~\cite{Li:2012wna} and single top quark~\cite{Aliev:2010zk,Kant:2014oha} events.
The $\ttbar$ samples are normalized to a cross section determined at NNLO in QCD that includes the resummation of the next-to-next-to-leading logarithmic (NNLL) soft-gluon terms~\cite{Beneke:2011mq,Cacciari:2011hy,Czakon:2011xx,Baernreuther:2012ws,Czakon:2012zr,Czakon:2012pz,Czakon:2013goa}.
Monte Carlo samples of other SM background processes are normalized to cross sections obtained from the MC event generators at either LO or NLO in QCD.
The SUSY cross sections are computed at approximately NNLO plus NNLL precision with all other SUSY particles assumed to be heavy and decoupled~\cite{Beenakker:1996ch,Kulesza:2008jb,Kulesza:2009kq,Beenakker:2009ha,Beenakker:2011fu,Borschensky:2014cia,Beenakker:2016lwe}.

To improve the modeling of the multiplicity of additional jets either from initial-state radiation (ISR) or final-state radiation (FSR),
simulated SM and SUSY events are reweighted so as to make the jet multiplicity agree with data.
The reweighting is applied to all SUSY samples but only to
2016 SM samples.
No reweighting is applied for 2017 and 2018 SM simulation because of the improved tuning of the MC generators mentioned above.
The procedure is based on a comparison of the light-flavor jet multiplicity in dilepton $\ttbar$ events in data and simulation.
The comparison is performed after selecting events with two leptons and two {\PQb}-tagged jets, which are jets identified as originating from the fragmentation of bottom quarks.
The reweighting factors obtained vary from 0.92 to 0.51 for one to six additional jets.
The uncertainties in the reweighting factors are evaluated as half of the deviation from unity.
These uncertainties cover the data-simulation differences observed in $\ttbar$ enriched validation samples obtained by selecting events with an $\Pe\PGm$ pair and at least one {\PQb}-tagged jet.

The \ptmiss and its vector ($\ptvecmiss$), defined in
Section~\ref{sec:evtsel}, are key ingredients of the analysis.
The modeling of their resolution in the simulation is studied in
\gjets samples for each data taking period.
Based on these studies, the simulated \ptmiss resolution is corrected with scale factors, the magnitudes of which are around 10\% for the 2018 data and up to 15\% for the latter subset of the 2017 data.
The correction factors for the earlier subset of the 2017 data, or the entire 2016 data are close to unity.
The variations seen in the \ptmiss resolution factors in the three data taking periods are mainly caused by different pileup and detector conditions, which are addressed in the next section.

\section{Event reconstruction and search strategy}
\label{sec:evtsel}

The overall strategy of the analysis follows that of the search presented in Ref.~\cite{Sirunyan:2017xse}.
Three categories of search regions are defined.
The ``standard selection'' is designed to be sensitive to the majority of the top squark
scenarios under consideration with
$\Delta m\left(\PSQt,\PSGczDo\right) > m_{\PQt}$.
In this paper we use the symbol $\Delta m(\mathrm{a},\mathrm{b})$ to indicate the mass difference
between particles $\mathrm{a}$ and $\mathrm{b}$, and $m_{\mathrm{a}}$ to denote the mass of $\mathrm{a}$.
Two additional sets of signal regions are used to target decays of the top squark to a top quark and a neutralino with mass splittings between these particles of either
$\Delta m\left(\PSQt,\PSGczDo\right)\sim m_{\PQt}$,
or $\Delta m\left(\PSQt,\PSGczDo\right)\sim m_{\PW}$.

\subsection{Event reconstruction}
\label{sec:evtreco}

The events used in this analysis are selected using triggers that require either large \ptmiss, or the presence of an isolated electron or muon.
The $\ptvecmiss$ is first computed from the negative vector
sum of the $\pt$ of all particle-flow candidates, described below. The trigger selects events with $\ptmiss > 120\GeV$.
The minimum requirement on the lepton $\pt$ varied between 27 and 35\GeV for electrons, and between 24 and 27\GeV for muons, depending on the data taking period.
The combined trigger efficiency, measured with a data sample of events with a large scalar sum of jet \pt, is greater than 99\% for events with $\ptmiss>250\GeV$ and lepton $\pt>20\GeV$.

The CMS event reconstruction is based on a particle-flow (PF) algorithm~\cite{Sirunyan:2017ulk}.
The algorithm combines information from all CMS subdetectors to identify charged and
neutral hadrons, photons, electrons, and muons, collectively referred to as PF candidates.

Each event must contain at least one reconstructed $\pp$ interaction vertex.
The reconstructed vertex with the largest value of the summed $\pt^2$ of physics objects is taken to be the primary vertex (PV).
The physics objects are the objects reconstructed by the anti-$\kt$ jet finding algorithm~\cite{Cacciari:2005hq,Cacciari:2008gp,Cacciari:2011ma}
with the tracks assigned to the vertex as inputs, and the associated missing transverse momentum (\mht), taken as the magnitude of the negative vector sum of the \pt of those jets.

Events with possible contributions from beam halo interactions or anomalous noise in the calorimeter are rejected using dedicated filters~\cite{Chatrchyan:2011tn}.
For the 2017 and 2018 data taking periods, the ratio of the scalar sums of jet \pt within $\abs{\eta}<5.0$ and of jet \pt within $\abs{\eta}<2.4$ is required to be smaller than 1.5 to reject events with significant \ptmiss arising from noise in the ECAL endcap forward region.
Additionally, during part of the 2018 data taking period, two sectors of the HCAL endcap detector experienced a power loss. The affected data sample size is about 39\fbinv. As the identification of both electrons and jets depends on correct energy fraction measurements, events from the affected data taking
periods containing an electron or a jet in the region $-2.4<\eta<-1.4$ and azimuthal angle $-1.6<\phi<-0.8$ radians are rejected.
The effect is estimated to be an approximately 2\% loss in signal and background acceptance for the full dataset.
The simulation is corrected to take this loss into account.

\begin{figure*}[htbp]
\centering
\includegraphics[width=0.49\textwidth]{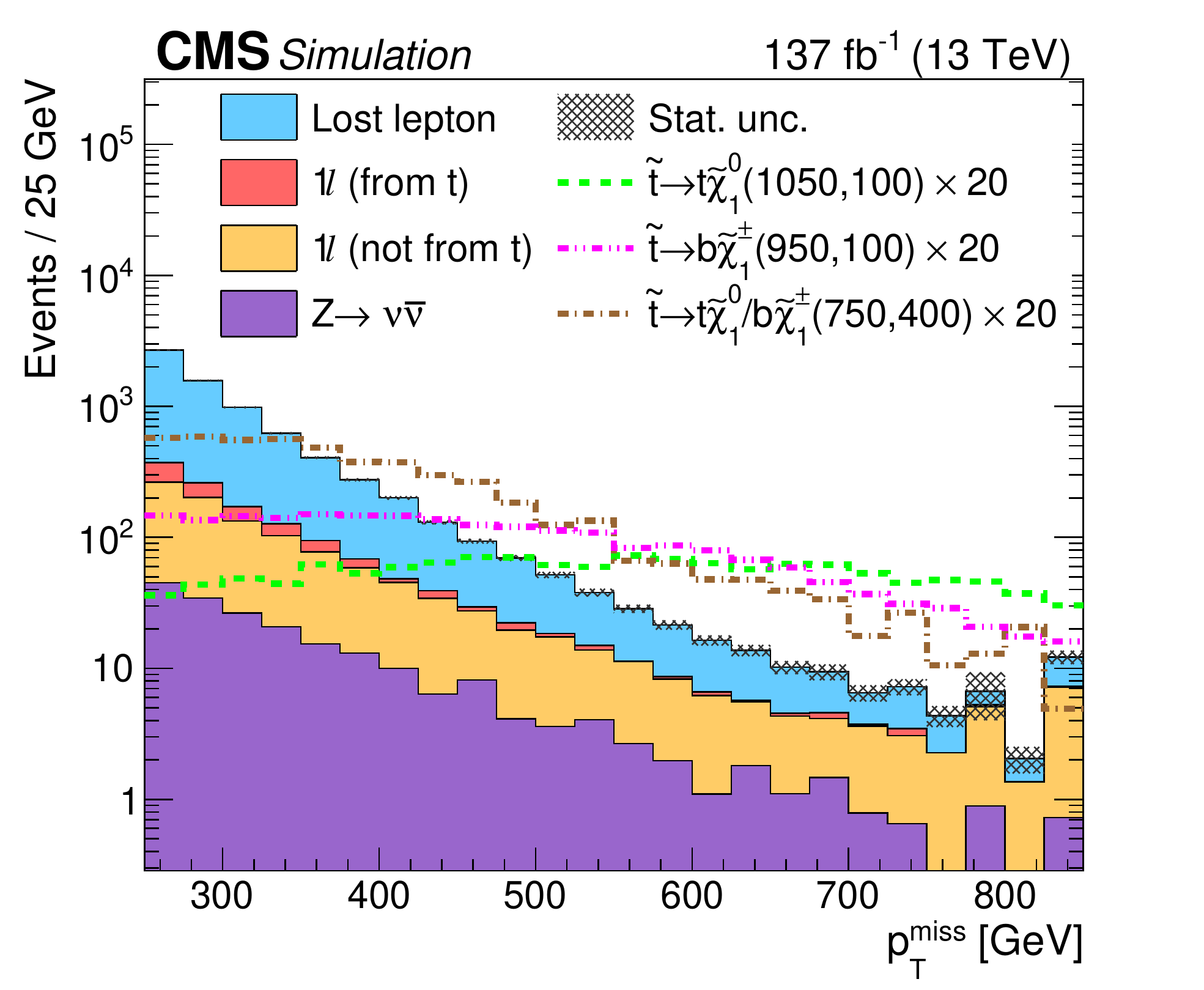}
\includegraphics[width=0.49\textwidth]{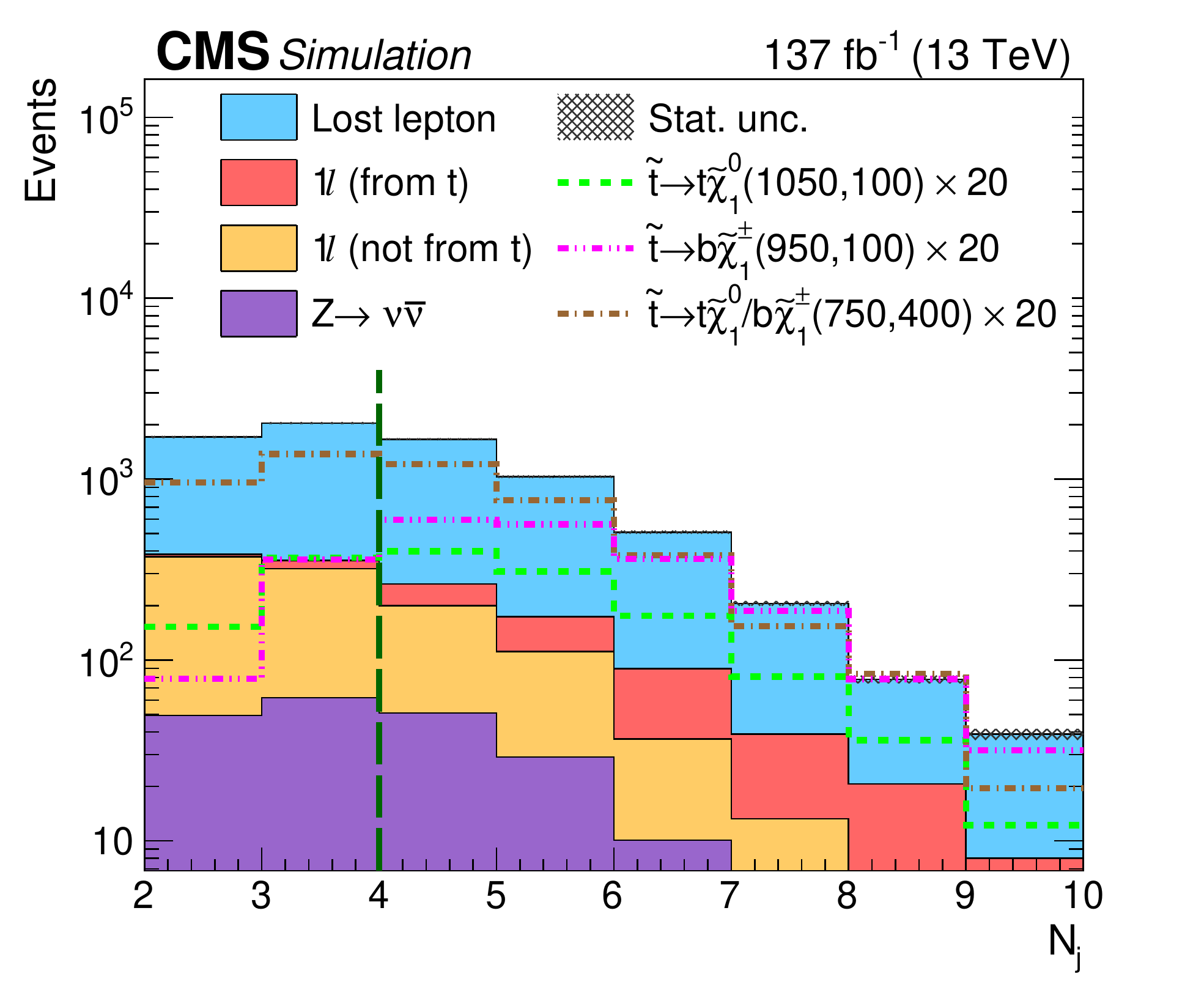} \\
\includegraphics[width=0.49\textwidth]{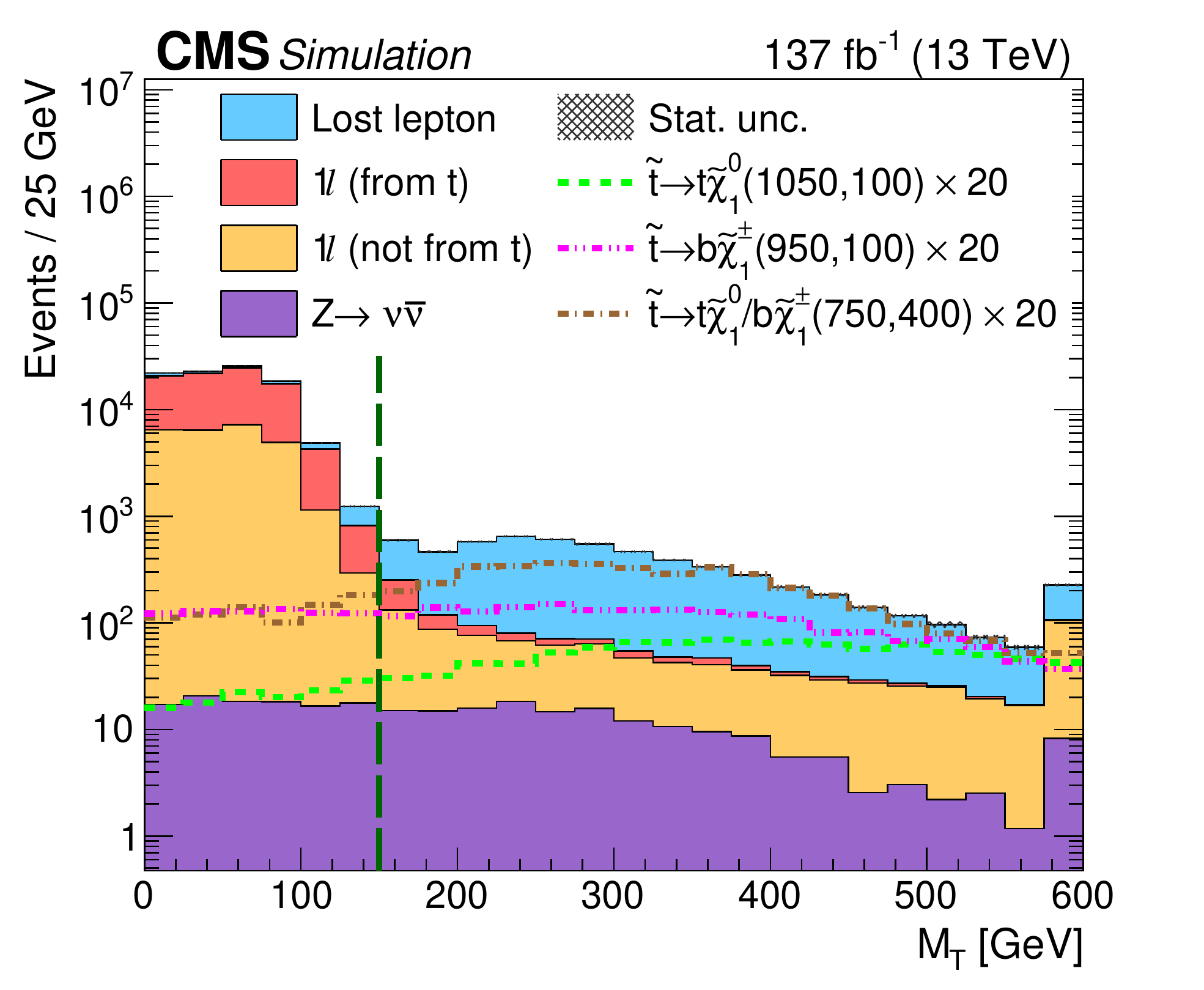}
\includegraphics[width=0.49\textwidth]{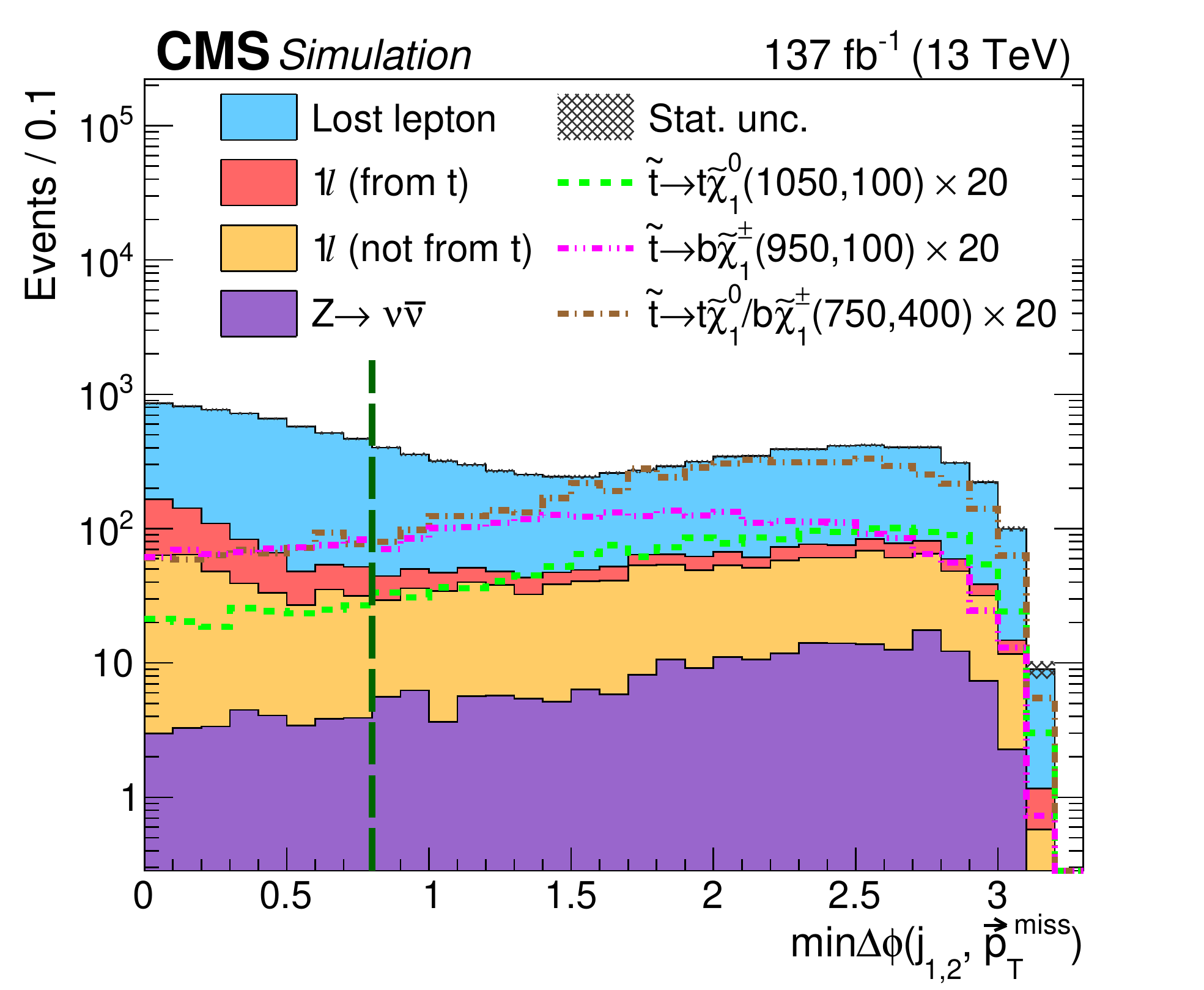}
\caption{
  The distributions of \ptmiss (upper left) and \NJ (upper right) are shown after applying the preselection requirements of Table~\ref{tab:sels}, including the requirement on the variable shown, and the distributions of \MT (lower left) and \minDPhiMETjet (lower right) are shown after applying the preselection requirements, excluding the requirement on the variable shown with the green, dashed vertical line marking the location of the requirement.
The stacked histograms for the SM background contributions (categorized as described in Section~\ref{Sec:BkgEst}) are from the simulation to illustrate the discriminating power of these variables. The gray hashed region indicates the statistical uncertainty of the simulated samples.
  The last bin in each distribution includes the overflow events. The expectations for three signal hypotheses are overlaid, and the corresponding numbers in parentheses in the legends refer to the masses of the top squark and neutralino, respectively. For models with {\PQb}$\chgo$ decays, the mass of the chargino is chosen to be $(m_{\PSQt} + m_{\lsp})/2$.
}
\label{fig:SRplots-cut}
\end{figure*}

After these initial requirements, we apply an event preselection summarized in Table~\ref{tab:sels}
and described below.
Selected events are required to have exactly one electron~\cite{Khachatryan:2015hwa} or muon~\cite{Sirunyan:2018fpa} originating from the PV and isolated from other activity in the event.
Leptons are identified as isolated if the scalar sum of the \pt of all PF candidates in a cone around the lepton, excluding the lepton itself, is less than 10\% of the lepton \pt.
Typical lepton selection efficiencies are approximately
85\% for electrons and 95\% for muons, depending on \pt and $\eta$.

\begin{table*}[bthp]
\centering
  \setlength{\extrarowheight}{.7em}
\topcaption{
Summary of the event preselection requirements.
The magnitude of the negative vector sum of the \pt of all jets and leptons in the event is denoted by \mht.
The symbols $\pt^{\ell}$ and $\eta^{\ell}$ correspond to the transverse momentum and pseudorapidity
of the lepton. The symbol $\ptsum$ is the scalar sum of the \pt of all (charged) PF candidates in a cone around
the lepton (track), excluding the lepton (track) itself. Finally, $\Nmedb$ and $\Nsoftb$ are the multiplicity
of {\PQb}-tagged jets (medium working point) and soft {\PQb} objects, respectively.
}
\label{tab:sels}
    \begin{tabular}{ ll }
      \hline
      \multirow{3}{*}{Trigger (2016)} &  $\ptmiss>170\GeV$ or \\
      & $\ptmiss>120\GeV$ and $\mht>120\GeV$ or \\
      & isolated $\PGm (\Pe)$ with $\pt^{\ell} > 24 (25)\GeV$ \\
      \multirow{2}{*}{Trigger (2017, 2018)} & $\ptmiss>120\GeV$ and $\mht>120\GeV$ or \\
      & isolated $\PGm (\Pe)$ with $\pt^{\ell} > 27 (35)\GeV$ \\
      [\cmsTabSkip]
      \multirow{2}{*}{$\ptsum$ cone size} & for $\PGm$ or $\Pe$: $\Delta R = \min[ \max(0.05, 10\GeV / \pt^\ell), 0.2 ]$ \\
      & for track: $\Delta R = 0.3$ \\
      [\cmsTabSkip]
      \multirow{2}{*}{Lepton} & $\PGm (\Pe)$ with $\pt^{\ell} > 20\GeV$, $\abs{\eta^{\ell}} <2.4~(1.44)$ \\
      & $\ptsum < 0.1 \times \pt^{\ell}$ \\
      [\cmsTabSkip]
      \multirow{2}{*}{Veto lepton} & $\PGm$ or $\Pe$ with $\pt^{\ell} > 5\GeV$, $\abs{\eta^{\ell}} <2.4$ \\
      & $\ptsum < 0.2 \times \pt^{\ell}$ \\
      \multirow{2}{*}{Veto track} & Charged PF candidate, $\pt > 10\GeV$, $\abs{\eta} <2.4$ \\
      & $\ptsum < \min\left(0.1 \times \pt, \mathrm{6}\GeV\right)$ \\
      [\cmsTabSkip]
      Jets & $\pt > 30\GeV$, $\abs{\eta} <2.4$, $\NJ\ge2$ \\
      \multirow{2}{*}{{\PQb} tagging} & $\Nmedb\ge1$ for standard and $\Delta m\left(\PSQt,\PSGczDo\right)\sim m_{\PQt}$ selection\\
      & $\Nsoftb\ge1$ for $\Delta m\left(\PSQt,\PSGczDo\right)\sim m_{\PW}$ selection \\
      [\cmsTabSkip]
      \ptmiss & $>250\GeV$ \\
      \MT & $>150\GeV$ \\
      \multirow{2}{*}{$\minDPhiMETjet$} & $>0.8$ radians for standard search \\
      & $>0.5$ radians for compressed scenarios \\
      \hline
    \end{tabular}
\end{table*}

The PF candidates are clustered into jets using the anti-\kt algorithm with a distance parameter of 0.4.
Jet energies are corrected for contributions from multiple interactions in the same or adjacent beam crossing (pileup)~\cite{cacciari-2008-659, jetid13TeV} and to account for nonuniformity in the detector response.
These jet energy corrections are propagated to the
calculation of $\ptvecmiss$~\cite{Khachatryan:2016kdb,CMS-DP-2018-028}.

Jets in the analysis are required to be within $\pt > 30\GeV$ and $\abs{\eta} < 2.4$, and
the number of these jets ($\NJ$) is required to be at least two.
Jets overlapping with the selected lepton within a cone radius of $\Delta R=0.4$ are not counted.
The distribution of the number of jets after the preselection requirements is shown in Fig.~\ref{fig:SRplots-cut} (upper right).
The jet multiplicity is used to define the signal region bins to optimize sensitivity for a variety of signal models and SUSY particle masses, as shown in this figure.

After these requirements, jets originating from a bottom quark fragmentation are identified as {\PQb}-tagged jets by the combined secondary vertex algorithm using a deep neural network (DeepCSV)~\cite{Sirunyan:2017ezt}.
The preselection requires at least one {\PQb}-tagged jet with either a medium or tight working point. The threshold on the discriminator value corresponding to the medium (tight) working point
is chosen so that the tagging rate for light-flavor jets is about 1\% (0.1\%), corresponding to an efficiency to identify a jet originating from a bottom-flavored hadron of 65--80 (40--65)\%, for jet \pt of 30--400\GeV.

To enhance sensitivity to signal scenarios with a compressed mass spectra, we use a secondary vertex (SV), not associated to jets or leptons, to identify soft {\PQb} hadrons~\cite{Sirunyan:2017wif} with $\pt > 1\GeV$ and $\abs{\eta} < 2.5$.
The SV is reconstructed by the inclusive vertex finding algorithm~\cite{Khachatryan:2011wq}.
At least two tracks must be associated to the SV and the sum of the transverse momenta of all the associated tracks is required to be below 20\GeV.
The distance between the SV and the PV must be ${<}3\unit{cm}$ and the significance of this distance is required to be $>$4. The cosine of the pointing angle defined by the scalar product between the distance vector, $\overrightarrow{(\text{PV,SV})}$, and
the $\vec{p}_{\text{SV}}$, where the $\vec{p}_{\text{SV}}$ is the total three-momentum of the tracks associated with the SV, must be $>$0.98.
These requirements help suppress background from light-flavor hadrons and jets.
Events containing objects that pass these selections, are said to contain a ``soft {\PQb} object''. These requirements result in a 40--55 (2--5)\% efficiency to select a soft {\PQb} object originating from a soft bottom-flavor (light-flavor) hadron.
As listed in Table~\ref{tab:sels},
the preselection requires the presence of at least one soft {\PQb} object in the signal regions dedicated to the compressed mass spectra.

The background processes relevant for this search are semileptonic or dileptonic \ttbar ($\ttbar\to 1\ell +\text{X}\xspace$ or $\ttbar\to 2\ell + \text{X}\xspace$), single top quark production
(mostly in the $\PQt\PW$ channel), \wjets, and processes containing a \PZ boson decaying into a pair of neutrinos ($\Znunu$), such as $\ttbar\PZ$ or $\PW\Z$.
Contributions to the background from semileptonic $\ttbar$ and \wjets are heavily suppressed by requiring
in the preselection that the transverse mass (\MT) be greater than 150\GeV and the \ptmiss to be greater than 250\GeV, as shown in Fig.~\ref{fig:SRplots-cut} (upper left and lower left, respectively).
The \MT is defined as $ \sqrt{\smash[b]{2\pt^\ell\ptmiss[1-\cos(\Delta\phi)]}}$
with $\pt^\ell$ denoting the lepton \pt, and $\Delta\phi$ the azimuthal separation between the lepton direction and \ptvecmiss.

In addition, to suppress background from processes with two leptonically decaying \PW bosons, primarily $\ttbar$ and $\PQt\PW$,
we also reject events containing either an additional lepton passing a loose
selection (denoted as ``veto lepton'' in Table~\ref{tab:sels})
or an isolated track.
Further rejection is achieved by requiring that the minimum angle in the transverse plane between the \ptvecmiss and the directions of the two leading \pt jets in the event (denoted as $j_{1,2}$), \minDPhiMETjet, is greater than 0.8 or 0.5, depending on the signal region.
This can be seen from the distribution of \minDPhiMETjet, after applying the rest of the preselection requirements, shown in Fig.~\ref{fig:SRplots-cut} (lower right).

In addition to the preselection requirements, we also use
two deep neural networks (DNNs) to categorize events based on the identification
of hadronically decaying top quarks.

One DNN, referred to as the resolved tagger, uses the DeepResolved algorithm to identify hadronically decaying top quarks with a moderate Lorentz boost. The decay products of these objects result in three separate jets (resolved top quark decay).
The DeepResolved algorithm identifies top quarks decaying into three distinct jets passing the selection requirements.
The three jets ($\pt > 40$, 30, 20\GeV) of each candidate must have an invariant mass between 100 and 250\GeV, no more than one of the jets can be identified
as a {\PQb}-tagged jet, and the three jets must all lie within a cone of $\Delta R < 3.14$ of the trijet centroid.

A neural network is used to distinguish trijet combinations which match to a top quark versus those which do not.
The network uses high-level information such as the invariant mass of the trijet system and of the individual dijet
pairs, as well as kinematic information from each jet. This includes its Lorentz vector, DeepCSV heavy-flavor discriminator
values, jet shape variables, and detector level particle multiplicity and energy fraction variables.
The network is trained using both \ttbar and QCD simulation, and data as training inputs. The simulation is used to define the examples of signal and background.
The signal is defined as any trijet passing the preselection requirements, where each jet is matched to a generator level daughter of a top quark within a cone of
$\Delta R < 0.4$ and the overall trijet system is matched to the generator level top quark within a cone of $\Delta R < 0.6$.
The background category is defined as any trijet combination that is not categorized as signal.
This includes trijet combinations for which some, but not all, of the jets match top decay products.
The data is included in the training to inhibit the network from learning features of the MC which are not present in data.
This is achieved through a technique called domain adaption via gradient reversal~\cite{ganin2014unsupervised}.
With this method, an additional output is added to the neural network to distinguishing between trijet candidates from QCD simulation and a QCD-enriched data sample.
The main network is then restricted to minimize its ability to discriminate simulation from data.
This yields a network with good separation between signal and background while minimizing over-fitting on features that exist only in simulation.
Before the final selection of trijets as top quarks can be made, any trijet candidates that may share the jets with another candidate must be removed. This is achieved by always favoring the candidate with a higher top discriminator value as determined by the neural network.
The reconstructed candidates are identified as hadronic tops when the neural network discriminator is above the threshold corresponding to an efficiency of 45$\%$ and the mistagging rate is 10$\%$ for dileptonic \ttbar events.

The second DNN, referred to as a merged tagger, uses the DeepAK8~\cite{CMS-PAS-JME-18-002} algorithm to identify top quarks with large boost,
where the decay products are merged into a single jet (merged top quark decay). The identification of this boosted top quark signature
is based on anti-\kt jets clustered with a distance parameter of 0.8.
The efficiency for lepton + hadronic-top events is 40\% and the mistagging rate is 5\% for dileptonic \ttbar events.

\subsection{Search strategy}
\label{sec:srdef}

\begin{figure*}[htbp]
\centering
\includegraphics[width=0.49\textwidth]{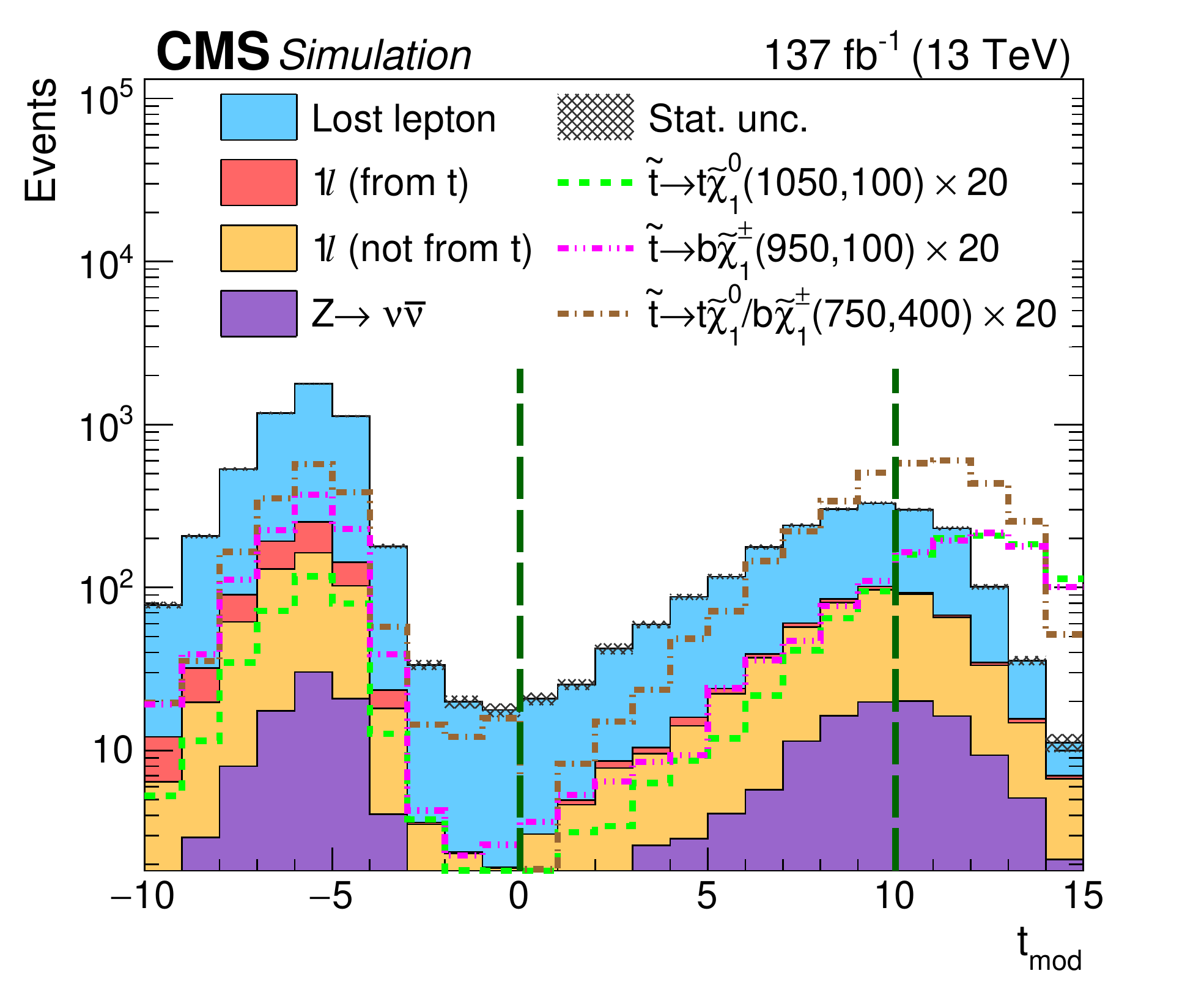}
\includegraphics[width=0.49\textwidth]{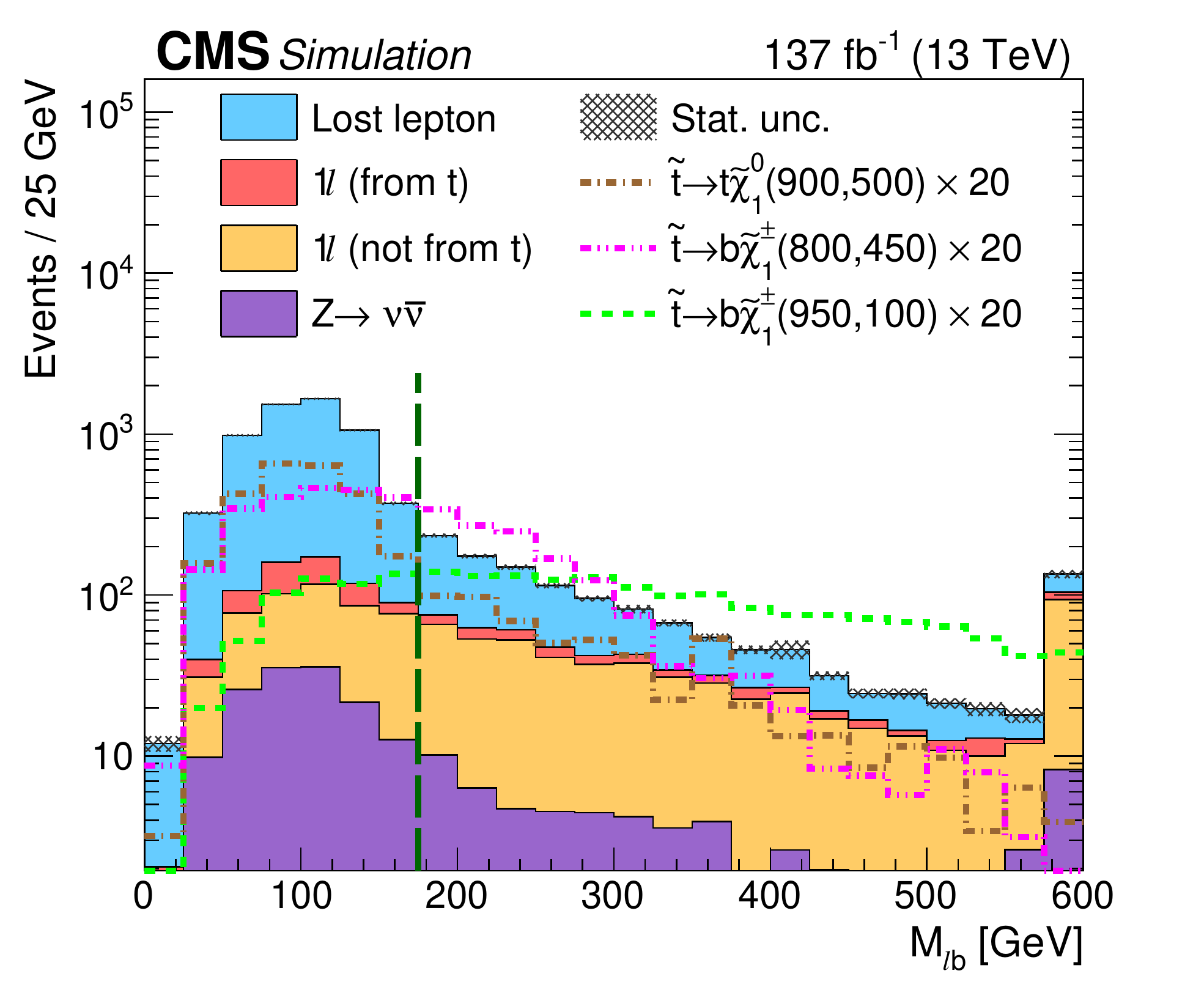}
\\
\includegraphics[width=0.49\textwidth]{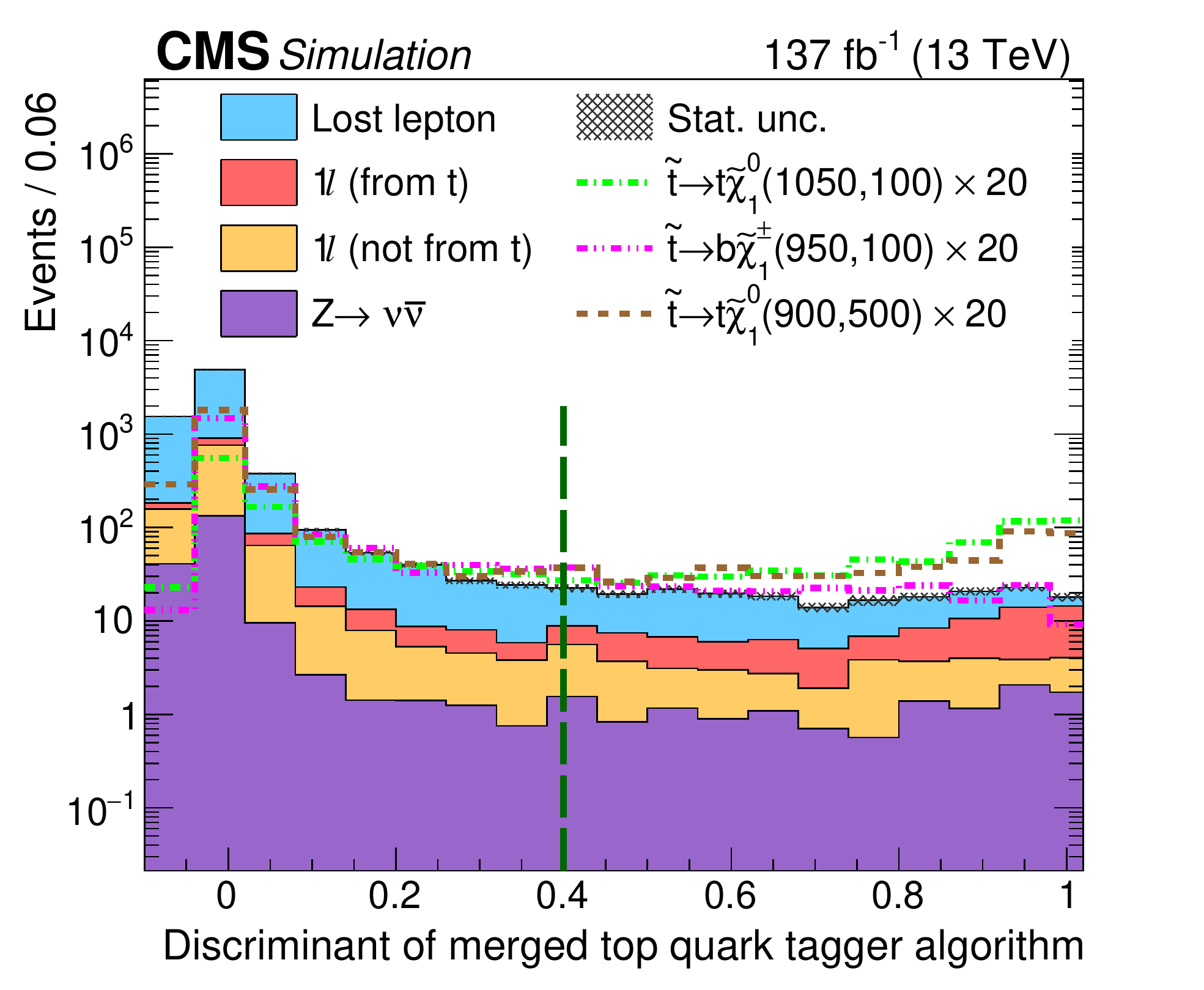}
\includegraphics[width=0.49\textwidth]{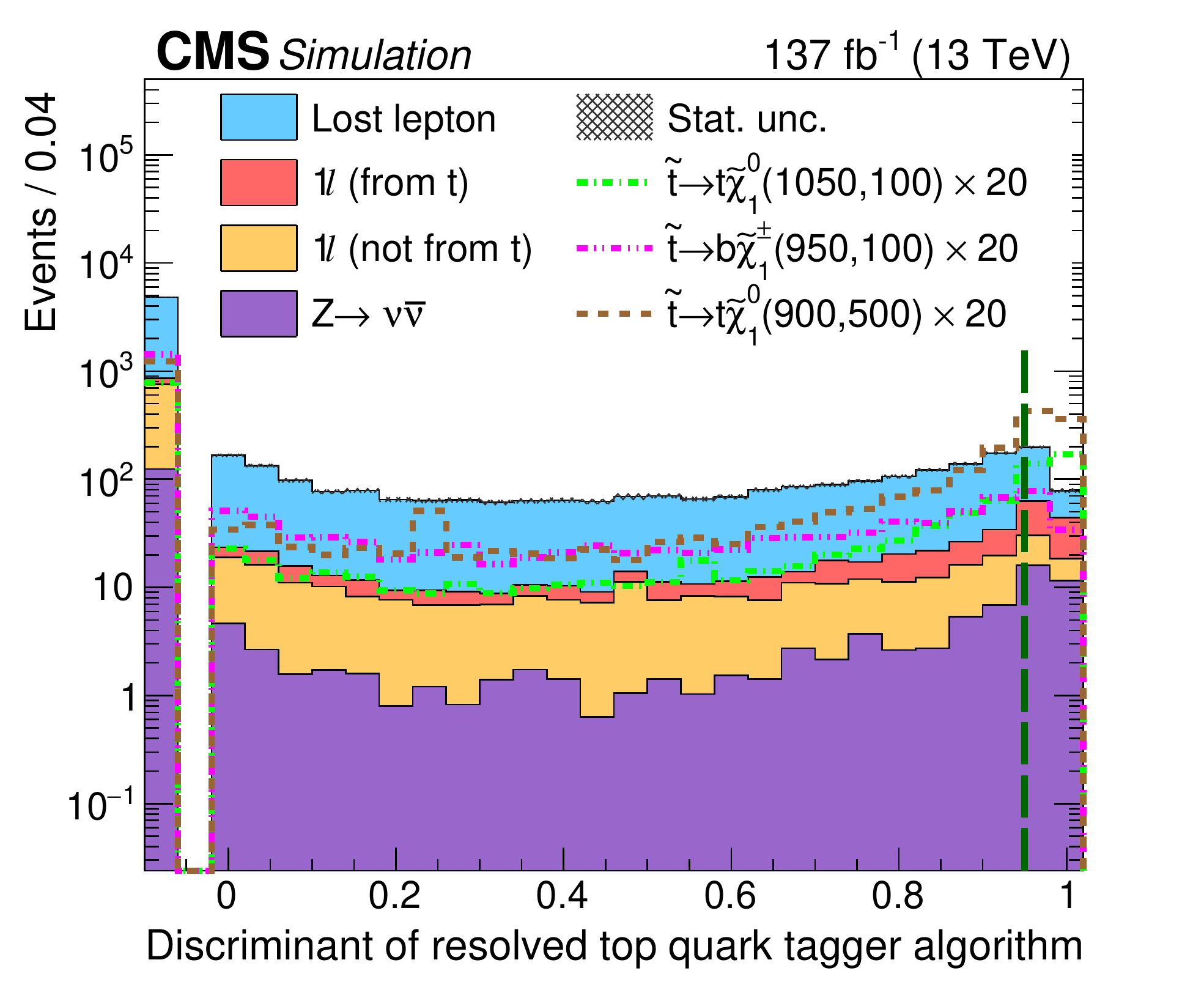}
\caption{
  The distributions of \tmod (upper left), \Mlb (upper right), the merged top quark tagging discriminant (lower left), and the resolved top quark tagging discriminant (lower right) are shown after the preselection requirements.
The green, dashed vertical lines mark the locations of the binning or tagging requirements.
The stacked histograms showing the SM background contributions (categorized as described in Section~\ref{Sec:BkgEst}) are from the simulation to illustrate the discriminating power of these variables. The gray hashed region indicates the statistical uncertainty of the simulated samples.
  Events outside the range of the distributions shown are included in the first or last bins. The expectations for three signal hypotheses are overlaid, and the corresponding numbers in parentheses in the legends refer to the masses of the top squark and neutralino, respectively.
 For models with {\PQb}$\chgo$ decays, the mass of the chargino is chosen to be $(m_{\PSQt} + m_{\lsp})/2$.
}
\label{fig:SRplots-var}
\end{figure*}

The signal regions for the standard search are summarized in Table~\ref{tab:SR}, and are defined by
categorizing events passing the preselection requirements based on $\NJ$, the number of identified hadronic top quarks,
\ptmiss, the invariant mass ($\Mlb$) of the lepton and the
closest {\PQb}-tagged jet in $\Delta R$,
and a modified version of the topness variable~\cite{Graesser:2012qy}, \tmod~\cite{Sirunyan:2016jpr}, which is defined as:
\begin{equation*}
\tmod = \ln(\min S),
\text{ with }
S = \frac{\left(m_{\PW}^2-(p_\nu+p_{\ell})^2\right)^2}{a_{\PW}^4} +
\frac{\left(m_{\PQt}^2 - (p_{\PQb}+p_{\PW})^2\right)^2}{a_{\PQt}^4},
\label{eq:tmod}
\end{equation*}
with resolution parameters $a_{\PW} = 5\GeV$ and $a_{\PQt} = 15\GeV$.
The \tmod variable is a $\chi^2$-like variable that discriminates signal from
leptonically decaying $\ttbar$ events: an event with a small value of \tmod is likely to be a dilepton $\ttbar$ event, while signal events tend to have
larger \tmod values.
The first term in its definition corresponds to the top quark decay containing the reconstructed lepton, and the second term corresponds to the top quark decay containing the missing lepton.
The $p_{\PW}$ in the second term symbolizes the momentum of the missing lepton and neutrino from the W decay.
The minimization of the variable $S$ is done with respect to all components of the three momentum $\vec{p}_{\PW}$
and the component of the three momentum $\vec{p}_\nu$ along the beam line with the constraints that $\ptvecmiss=\vec{p}_{\mathrm{T},\PW}+\vec{p}_{\mathrm{T},\nu}$ and $p^2_{\PW}=m^2_{\PW}$.
The distribution of \tmod for events passing the preselection is shown in Fig.~\ref{fig:SRplots-var} (upper left).
The \tmod distribution is split into three bins, each sensitive to a different mass splitting of the top squark and neutralino.

\begin{table*}[tbhp]
\centering
\topcaption{\protect The 39 signal regions of the standard selection, with
  each neighboring pair of values in the \ptmiss bins column defines a single signal region.
  At least one {\PQb}-tagged jet selected using the medium (tight)
  working point is required for search regions with $\Mlb$ lower (higher) than 175\GeV.
  For the top quark tagging categories, we use the abbreviations U for untagged, M for merged, and R
  for resolved.}
\label{tab:SR}
\begin{tabular}{c ccccl}
\hline
\multirow{2}{*}{Label} & \multirow{2}{*}{$\NJ$} & \multirow{2}{*}{\tmod} & \Mlb  & \cPqt tagging & \multirow{2}{*}{\ptmiss bins [{\GeVns}]} \\
                       &                        &                        & [{\GeVns}] & category    &   \\
\hline
A0 & \multirow{3}{*}{2--3}     & \multirow{3}{*}{$>$10}    & \multirow{3}{*}{$\le$175}  & \NA & [600, 750, $+\infty$]                \\
A1 &                           &                           &                            &  U & [350, 450, 600]                      \\
A2 &                           &                           &                            &  M & [250, 600]                           \\ [\cmsTabSkip]
B  &                 2--3      &                 $>$10     &              $>$175        & \NA & [250, 450, 700, $+\infty$]           \\ [\cmsTabSkip]
C  &                 $\geq$4   &                 $\le$0    &             $\le$175       & \NA & [350, 450, 550, 650, 800, $+\infty$] \\ [\cmsTabSkip]
D  &                 $\geq$4   &                 $\le$0    &              $>$175        & \NA & [250, 350, 450, 600, $+\infty$]      \\ [\cmsTabSkip]
E0 & \multirow{4}{*}{$\geq$4}  & \multirow{4}{*}{0--10}    & \multirow{4}{*}{$\le$175}  & \NA & [450, 600, $+\infty$]                \\
E1 &                           &                           &                            &  U & [250, 350, 450]                      \\
E2 &                           &                           &                            &  M & [250, 350, 450]                      \\
E3 &                           &                           &                            &  R & [250, 350, 450]                      \\ [\cmsTabSkip]
F  &                 $\geq$4   &                 0--10     &               $>$175       & \NA & [250, 350, 450, $+\infty$]           \\ [\cmsTabSkip]
G0 & \multirow{4}{*}{$\geq$4}  & \multirow{4}{*}{$>$10}    & \multirow{4}{*}{$\le$175}  & \NA & [450, 550, 750, $+\infty$]           \\
G1 &                           &                           &                            &  U & [250, 350, 450]                      \\
G2 &                           &                           &                            &  M & [250, 350, 450]                      \\
G3 &                           &                           &                            &  R & [250, 350, 450]                      \\ [\cmsTabSkip]
H  &                 $\geq$4   &                 $>$10     &               $>$175       & \NA & [250, 500, $+\infty$]                \\ \hline
\end{tabular}
\end{table*}

In events containing a leptonically decaying top quark,
the invariant mass of the lepton and the bottom quark jet from the same top quark decay is bound by
\begin{equation*}
\Mlb \leq m_{\PQt}\sqrt{1-\frac{m^{2}_{\PW}}{m^{2}_{\PQt} } }.
\label{eq:Mlbtheory}
\end{equation*}
This bound does not apply to either $\wjets$ events or signal
events, where the top squark decays to a bottom quark and a chargino.
To maintain acceptance to a broad range of signal scenarios, rather than requiring a selection on \Mlb, events are placed into low- or high-\Mlb categories if the value of \Mlb is less or greater than 175\GeV, respectively.
In signal regions with $\Mlb>175\GeV$, at least one jet is required to satisfy the tight {\PQb} tagging working point of the DeepCSV discriminator to suppress the background from $\wjets$ events.
The distribution of \Mlb in the signal regions is shown in Fig.~\ref{fig:SRplots-var} (upper right).
As seen from this figure, the low \Mlb regions are more sensitive to {\PQt}$\lsp$ and the $\Mlb>175\GeV$ are more sensitive to  {\PQb}{\chgo}.

Hadronic top quark taggers are used in signal regions sensitive to SUSY scenarios with hadronically decaying top quarks when most of the expected SM
background does not contain such a top quark decay. Therefore, the hadronic top taggers are deployed in the low
\Mlb, $\tmod\geq0$, and relatively modest \ptmiss signal regions.
Events containing two or three jets and $\ptmiss \le 600\GeV$, or at least four jets and $\ptmiss \le 450\GeV$, are categorized according to the presence of a merged top quark tag.
The resolved top quark tagger is used to further categorize events with four or more jets.
If an event contains both merged and resolved top quark tags, it is
placed in the \textit{merged} top category,
while events containing neither
are categorized as \textit{untagged}.
Distributions of the discriminant of the merged and resolved top quark taggers in the signal regions are also shown in Fig.~\ref{fig:SRplots-var} (lower left and lower right, respectively).

The small mass splitting in SUSY models with a compressed mass spectrum
results in soft decay products.
High values of
\ptmiss can only be caused by large boost from ISR.
As a result, in signal regions targeting these models
the jet with the highest \pt is expected to be from ISR and therefore
it is required to not be identified as a bottom quark jet.
We also impose an upper bound on the lepton \pt relative to the \ptmiss, since this requirement
provides
an additional handle to reject SM $\wjets$ and $\ttbar$ backgrounds.
Regions targeting signal scenarios with $\Delta m\left(\PSQt,\PSGczDo\right)\sim m_{\PQt}$ require at least five jets and at least one {\PQb}-tagged jet based on the DeepCSV medium working point.
For signal scenarios with $\Delta m\left(\PSQt,\PSGczDo\right)\sim m_{\PW}$, the bottom quarks are expected to have low \pt.
Therefore, in these regions the \NJ selection is relaxed to $\NJ \geq 3$ and instead of requiring the presence of a {\PQb}-tagged jet
we require the presence of a soft {\PQb} object.
Note that soft {\PQb} objects are included in the jet count in these regions.
The requirements for the two sets of signal regions targeting compressed mass spectrum SUSY scenarios are summarized in Table~\ref{tab:compSR}.

\begin{table*}[bthp]
\setlength{\extrarowheight}{.7em}
\topcaption{
Definitions of the total 10 search regions targeting signal scenarios with a compressed mass spectrum. Search regions for $\Delta m\left (\PSQt,\PSGczDo\right)\sim m_{\PQt}$
and $\sim m_{\PW}$ scenarios are labeled with the letter I and J, respectively. The symbol $\pt^{\ell}$ denotes the transverse momentum of the lepton.
Each neighboring pair of values in the \ptmiss bins column defines a single signal region.
}
\label{tab:compSR}
\centering
\begin{tabular}{ l l lllll }
\hline
\multicolumn{7}{c}{Compressed spectra with $\Delta m\left(\PSQt,\PSGczDo\right)\sim m_{\PQt}$} \\
\hline
Label I & \multirow{2}{*}{Selection criteria}
 & \multicolumn{4}{l}{$\NJ \geq5$, leading-$\pt$ jet not {\PQb}-tagged, $\Nmedb\ge1$, } & \\
 & & \multicolumn{3}{l}{$\pt^{\ell} < \max\left(50,\ 250 - 100 \times \Delta\phi(\ptvecmiss,\ptvec^{\ell})\right)\GeV$,} & & \\
 & \ptmiss bins [{\GeVns}] & \multicolumn{5}{l}{[250, 350, 450, 550, 750, $+\infty$]} \\
\hline
\multicolumn{7}{c}{Compressed spectra with $\Delta m\left(\PSQt,\PSGczDo\right)\sim m_{\PW}$} \\
\hline
Label J & \multirow{2}{*}{Selection criteria}
 & \multicolumn{4}{l}{$\NJ \geq3$, leading-$\pt$ jet not {\PQb}-tagged, $\Nsoftb\ge1$, } & \\
 & & \multicolumn{3}{l}{$\pt^{\ell} < \max\left(50,\ 250 - 100 \times \Delta\phi(\ptvecmiss,\ptvec^{\ell})\right)\GeV$,}  & &\\
 & \ptmiss bins [{\GeVns}] & \multicolumn{5}{l}{[250, 350, 450, 550, 750, $+\infty$]} \\
\hline
\end{tabular}
\end{table*}

\section{Background estimation}
\label{Sec:BkgEst}

Three categories of SM backgrounds remain after the selection requirements described in Section~\ref{sec:evtsel}.

\begin{itemize}
\item The lost-lepton background consists of events with two $\PW$ bosons decaying leptonically, where one of the leptons is either not reconstructed, or not identified. This background arises primarily from $\ttbar$ events, with a smaller contribution from single top quark processes. It is the dominant background in regions with low values of \Mlb, no top quark tag, or $\NJ\geq5$.  This background is estimated using a dilepton control sample.
\item The one-lepton background consists of events with a single $\PW$ boson decaying leptonically and without any additional source of genuine \ptmiss. The requirements of $\ptmiss>250\GeV$ and $\MT>150\GeV$ heavily suppress this background.
The one-lepton background is estimated from simulation when it originates from top quark decays (mainly semi-leptonic \ttbar).
Background events not originating from top quark decays, instead mainly from direct \PW production, are estimated using a control sample of events with no
\PQb-tagged jets.
\item The \Znunu background consists of events with a single leptonically decaying $\PW$ boson and a $\PZ$ boson that decays to a pair of neutrinos, \ie, $\pp \to \ttbar\PZ$ or $\PW\PZ$. This background is estimated using simulation.
\end{itemize}

\subsection{Lost-lepton background}
\label{sec:dilepton}

The lost-lepton background in each of the signal regions is estimated from corresponding dilepton control samples.
Each dilepton control sample is obtained with the signal selections except for the requirement of a second isolated lepton with $\pt>10\GeV$ and the removal of the lepton, track, and tau vetoes.
The estimated background in each search region is obtained from the yield of data events in the corresponding
control sample and a transfer factor obtained from simulation, $R^{\text{lost-}\ell / 2\ell}_{\text{MC}}$.
The transfer factor is defined as the ratio of the expected lost-lepton yield in the signal region and the yield of dilepton SM events in the control sample.
These transfer factors are validated by checking the modeling of lepton reconstruction and selections as well as the kinematical properties of leptons in simulation.
Corrections obtained from studies of samples of $\cPZ,\,\JPsi \to \ell \ell$ events are applied to the transfer factor to account for
differences in lepton reconstruction and selection efficiencies between data and simulation.
The kinematical properties of leptons are well modeled in simulation and have a data to simulation agreement within 10$\%$ or better.
Simulation shows that the dilepton control sample have high purity (70--80\%) of the main processes (dileptonic \ttbar and $\PQt\PW$) contributing to the lost-lepton background.
Small contamination from semileptonic \ttbar and other process, where the additional lepton is a fake or non-prompt lepton, are subtracted from the control sample data yields.

\begin{figure*}[thb]
\centering
\includegraphics[width=0.49\textwidth]{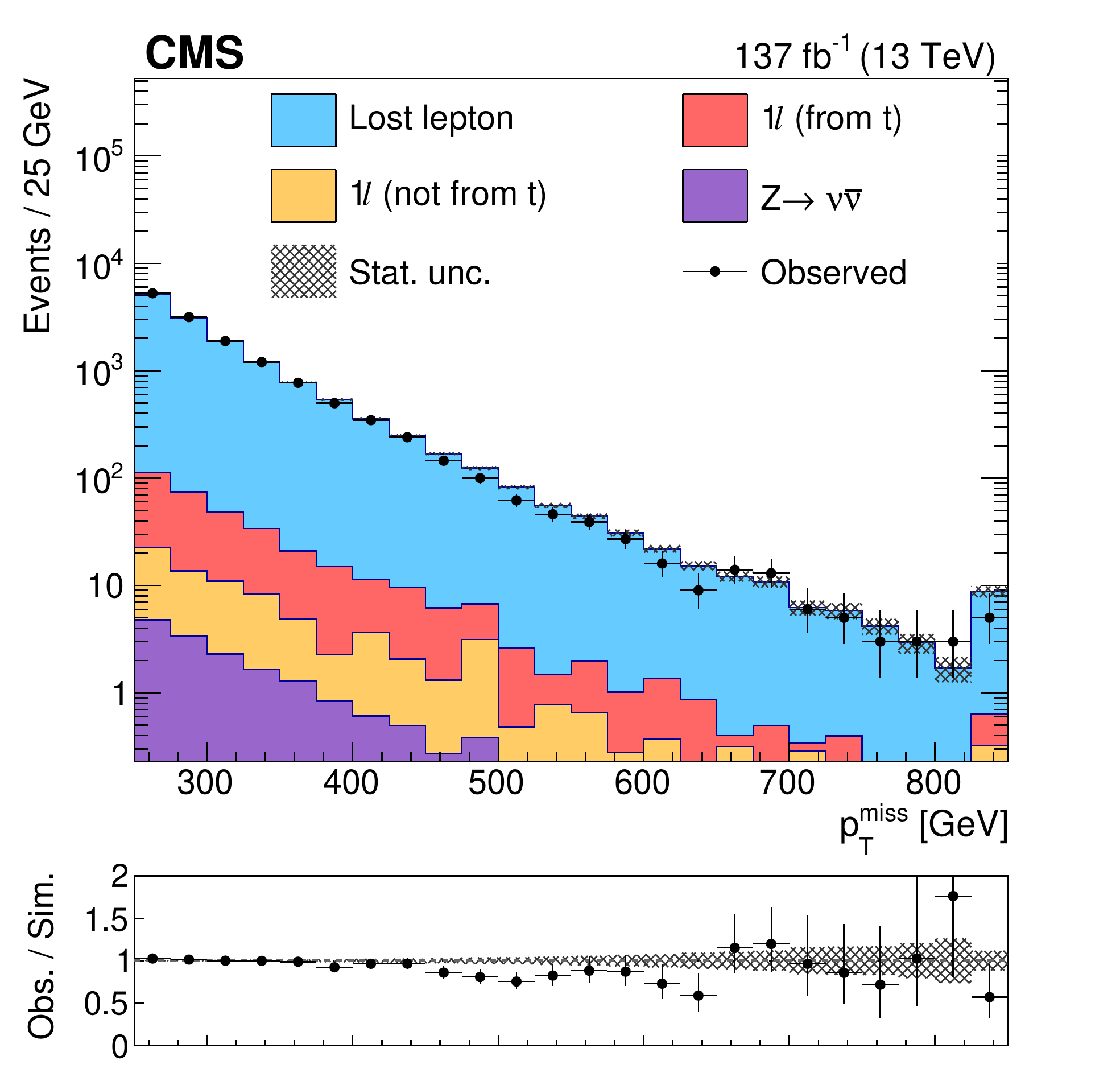}
\includegraphics[width=0.49\textwidth]{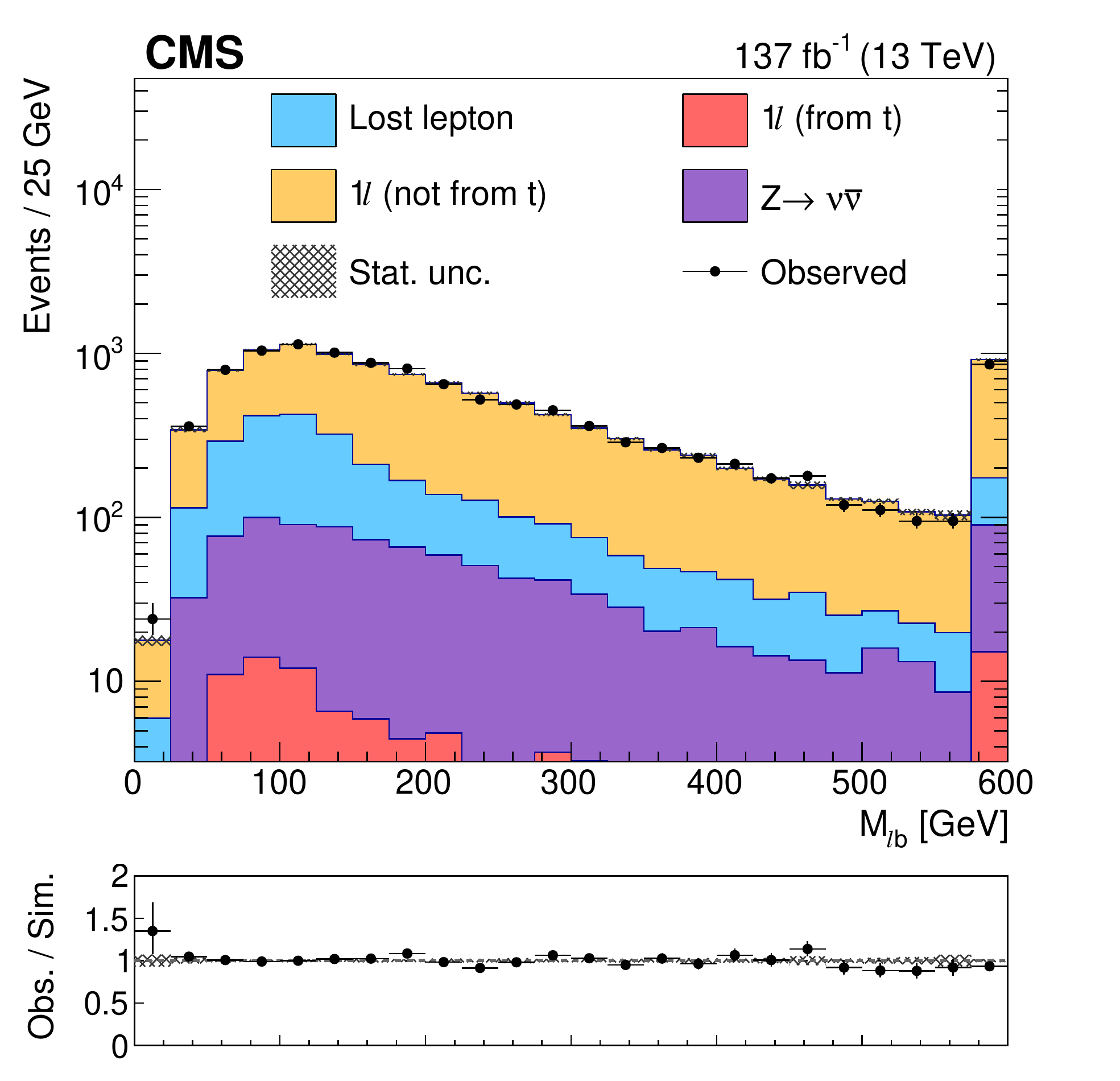}
\caption{
  Distributions of kinematic variables in the
  inclusive control samples used for the background estimation.
  The gray hashed region indicates the statistical uncertainty of the simulated samples.
  The distributions for data are shown as points with error bars corresponding to the statistical uncertainty.
  The stacked histograms show the expected SM background contributions from simulation, normalized to the number of events observed in data.
  The last bin in each distribution also includes the overflow.
  Left: Distribution of \ptmiss in the dilepton control sample.
  Right: Distribution of \Mlb in the 0{\PQb} control sample.
}
\label{fig:CRplots}
\end{figure*}

When defining the \ptmiss in this control sample, the trailing lepton \ptvec is added to \ptvecmiss to enhanced data statistics and all \ptvecmiss related quantities are recalculated.
The distribution of \ptmiss for after this addition is shown in Fig.~\ref{fig:CRplots} (left) for an inclusive selection.

Some control samples only contain a small number of events.
These samples, corresponding to multiple \ptmiss bins, are combined into a single control sample until the expected yield in simulation is at least five events, as detailed in Table~\ref{tab:metextrCR2L}.
The number of data events in the combined control sample is used
to estimate the sum of expected background
events in the corresponding signal regions. This sum is then
distributed across \ptmiss bins according to the expectation from simulation using an extrapolation factor $k(\ptmiss)$.
Additional corrections to account for the \ptmiss shape mismodeling observed in simulation with respect to data are derived in an orthogonal \ttbar enriched dilepton sample and applied to the simulation in these regions.

\begin{table*}[hbt]
\topcaption{
  Dilepton control samples that are combined when estimating the lost-lepton background.}
\label{tab:metextrCR2L}
\centering
\begin{tabular}{c lll l}
  \hline
 Label & \multicolumn{3}{c}{Selection}& {\ptmiss bins [GeV]} \\
  \hline
 A0 & 2--3 jets,    & $\tmod > 10$,     & $\Mlb \le 175\GeV$  & [600, 750, $+\infty$] \\
 B  & 2--3 jets,    & $\tmod > 10$,     & $\Mlb > 175\GeV$    & [450, 700, $+\infty$] \\
 C  & $\geq$4 jets, & $\tmod \le 0$,    & $\Mlb \le 175\GeV$  & [650, 800, $+\infty$] \\
 E0 & $\geq$4 jets, & $0< \tmod\le 10$, & $\Mlb \le 175\GeV$  & [450, 600, $+\infty$] \\
 G0 & $\geq$4 jets, & $\tmod > 10$,     & $\Mlb \le 175\GeV$  & [550, 750, $+\infty$] \\
 H  & $\geq$4 jets, & $\tmod > 10$,     & $\Mlb > 175\GeV$    & [250, 500, $+\infty$] \\ [\cmsTabSkip]
 I  & $\geq$5 jets, & $\Nmedb \ge 1$,   & $\Nsoftb \ge 0$      & [550, 750, $+\infty$] \\
 J  & $\geq$3 jets, & $\Nmedb \ge 0$,   & $\Nsoftb \ge 1$      & [550, 750, $+\infty$] \\
  \hline
\end{tabular}
\end{table*}

The lost-lepton background in each signal region, $N^{\text{SR}}_{\text{lost-}\ell}$, is obtained by scaling the number of events in the control sample, $N^{\text{CR}}_{2\ell}$, using the transfer factor $R^{\text{lost-}\ell / 2\ell}_{\text{MC}}$ and the \ptmiss extrapolation factor $k(\ptmiss)$ as follows:
\begin{equation} \label{dilepton estimate}
N^{\text{SR}}_{\text{lost-}\ell} = N^{\text{CR}}_{2\ell}\, R^{\text{lost-}\ell / 2\ell}_{\text{MC}}\, k(\ptmiss).
\end{equation}

The dominant uncertainties in the transfer factors are the statistical uncertainties in the simulated samples,
the uncertainties in the lepton efficiencies, and the uncertainties in the jet energy scale.
These uncertainties range between 3--68\%, 2--20\%, and 1--16\%, respectively.
Uncertainties in the {\PQb} tagging efficiency and in the choices of the renormalization and factorization scales are small.
The total uncertainty in the transfer factor is 6--100\%, depending on the region.
The uncertainty in the transfer factor is typically comparable to the statistical uncertainty in the control sample yield.
Associated uncertainties in the $k(\ptmiss)$ extrapolation factor used in the regions shown in Table~\ref{tab:metextrCR2L} were derived from an orthogonal \ttbar enriched dilepton sample.
The leading uncertainty associated with the \ptmiss extrapolation is the statistical uncertainty in the simulated samples (5--60\%).

\subsection{One-lepton background}
\label{sec:onelepton}

The one-lepton ($1\ell$)
background is suppressed by the
$\ptmiss>250\GeV$ and $\MT>150\GeV$ requirements.
This suppression is more effective for events with a $\PW$ boson originating from a top quark decay than for direct $\PW$ boson production ($\wjets$).
In the case of a top quark decay, the mass of the top quark sets bound at the mass of the lepton-neutrino system.
As a result, the contribution of semileptonic $\ttbar$ events to the tail of the \MT distribution is caused by \ptmiss resolution effects, while in
the case of $\wjets$ events the contribution from off-shell $\PW$ bosons is dominant.

The semileptonic $\ttbar$ background is taken from simulation.
Studies with simulated samples indicate that the contribution to the total background from semileptonic $\ttbar$ events is less than 10\% in most search regions, except in a few regions with $\geq$1 top quark tags, where the contribution becomes as large as 30\% ~\cite{Sirunyan:2017xse}.
An uncertainty of 100$\%$ is assigned to cover the impact of the uncertainties in the \ptmiss resolution as measured in a photon data sample.

The $\wjets$ background is estimated from a control sample with no {\PQb}-tagged jets nor soft {\PQb} objects (0{\PQb} sample)
obtained by inverting the {\PQb}-tagging requirement.
Figure~\ref{fig:CRplots} (right) shows the \Mlb distribution in the 0{\PQb} control sample, where this quantity is computed from the jet with the highest value of the DeepCSV discriminant.
The modeling of this distribution in simulation is validated by comparing simulation and data in a $\wjets$ enriched control sample obtained by selecting events with 1--2 jets and $60 < \MT < 120\GeV$.

The $\wjets$ background estimate in each search region is obtained from the yield in the corresponding control samples and a transfer factor determined from simulation.
These control samples are shown to have high purity (70--80\%) of the $\wjets$ process in places where this background is more significant in the corresponding ($\Mlb > 175\GeV$) search region.
In other cases, the purity can go down to 50\%. Contamination from lost-lepton and other processes are subtracted from the control sample data yields.
The transfer factor, defined as the ratio of the expected one lepton (not from {\PQt}) yield in the signal region and the yield of events in the 0{\PQb} control sample, accounts for the acceptance and the {\PQb} tagging efficiency.
The transfer factors are validated by checking the differences in performance of the b tagging algorithm and the off-shell W production modeling between data and simulation. 
Corrections are applied for differences in {\PQb} tagging efficiencies between data and simulation.
The $\wjets$ kinematic properties in the 0{\PQb} control sample show good agreement between data and simulation as shown in Figure~\ref{fig:CRplots}.
As in the case of the lost-lepton background estimate, multiple control samples are combined into a single control sample
until the expected yield in simulation is at least five events, as detailed in Table~\ref{tab:metextrCR0b}. 

The dominant uncertainties in the transfer factors are the statistical uncertainties in the simulated samples,
the uncertainties in the {\PQb} tagging efficiencies, and the $\PW+\PQb(\cPaqb)$ cross section.  

\begin{table*}[hbt]
\topcaption{
Search regions where the corresponding 0{\PQb} control samples are combined when estimating the $\wjets$ background.}
\label{tab:metextrCR0b}
\centering
\begin{tabular}{c lll ll}
  \hline
  Label & \multicolumn{3}{c}{Selection}& \multicolumn{2}{l}{\ptmiss bins [{\GeVns}]} \\
  \hline
  C  & $\geq$4 jets, & $\tmod \le 0$,     & $\Mlb \le 175\GeV$ & [650, 800, $+\infty$] \\
  E0 & $\geq$4 jets, & $0< \tmod \le 10$, & $\Mlb \le 175\GeV$ & [450, 600, $+\infty$] \\
  G0 & $\geq$4 jets, & $\tmod > 10$,      & $\Mlb \le 175\GeV$ & [550, 750, $+\infty$] \\
  \hline
\end{tabular}
\end{table*}

\subsection{Background from events containing \texorpdfstring{\Znunu}{Z to invisible}}

The third category arises from $\ttbar\PZ$, $\PW\PZ$, and other rare multiboson processes. In all these processes, events from a leptonically decaying $\PW$ boson, and one or more $\PZ$ bosons decaying to neutrinos, enter the search regions.
In most search regions, $\ttbar\PZ$ is the most important process
contributing to this category. These backgrounds are estimated from simulation.
The contribution from $\ttbar\PZ$ is
normalized using the measured value
of the cross section~\cite{Sirunyan:2017uzs}.
This normalization results in a rescaling of the theoretical cross section by $1.17^{+0.10}_{-0.09}$, where the uncertainty is taken from
the statistical uncertainty in the measurement.

\section{Systematic uncertainties}
\label{sec:syst}

The contributions to the total uncertainty in the estimated backgrounds and expected signal yields are summarized in Table~\ref{tab:syst}.
The total uncertainty is generally larger at higher \ptmiss or when yields in the control samples become small.
Out of the uncertainties quoted, the theoretical uncertainties are correlated across the different data-taking periods because they are independent of the data-taking period. The uncertainties on lepton efficiency are also assumed to be fully correlated, but other experimental uncertainties are taken as uncorrelated between the different data-taking years.

\begin{table}[thb]
\centering
\topcaption{Summary of major systematic uncertainties. The range of values reflect their impact on the estimated backgrounds and signal yields in different signal regions. A 100\% uncertainty is assigned to the $1\ell$ (from {\PQt}) background estimated from simulation. }
\begin{tabular}{ lcccc }
\hline
Source                           & Signal  & Lost lepton   & $1\ell$ (not from {\PQt})  & \Znunu \\
\hline
Data statistical uncertainty         & \NA        & 5--50\% & 4--30\%  & \NA        \\
Simulation statistical uncertainty   & 6--36\%    & 3--68\% & 5--70\%  & 4--41\%    \\
\ttbar \ptmiss modeling              & \NA        & 3--50\% & \NA      & \NA        \\
Signal \ptmiss modeling              & 1--25\%    & \NA     & \NA      & \NA        \\
QCD scales                           & 1--5\%     & 0--3\%  & 2--5\%   & 1--40\%    \\
Parton distribution                  & \NA        & 0--4\%  & 1--8\%   & 1--12\%    \\
Pileup                               & 1--5\%     & 1--8\%  & 0--5\%   & 0--7\%     \\
Luminosity                           & 2.3--2.5\% & \NA     & \NA      & 2.3--2.5\% \\
$\PW+\PQb(\cPaqb)$ cross section    & \NA        & \NA     & 20--40\% & \NA        \\
$\ttbar\PZ$ cross section            & \NA        & \NA     & \NA      & 5--10\%    \\
System recoil (ISR)                  & 1--13\%    & 0--3\%  & \NA      & \NA        \\
Jet energy scale                     & 2--24\%    & 1--16\% & 1--34\%  & 1--28\%    \\
\ptmiss resolution                   & \NA        & 1--10\% & 1--5\%   & \NA        \\
Trigger                              & 2--3\%     & 1--3\%  & \NA      & 2--3\%     \\
Lepton efficiency                    & 3--4\%     & 2--12\% & \NA      & 1--2\%     \\
Merged {\cPqt} tagging efficiency    & 3--6\%     & \NA     & \NA      & 5--10\%    \\
Resolved {\cPqt} tagging efficiency  & 5--6\%     & \NA     & \NA      & 3--5\%     \\
{\PQb} tagging efficiency           & 0--2\%     & 0--1\%  & 1--7\%   & 1--10\%    \\
Soft {\PQb} tagging efficiency      & 2--3\%     & 0--1\%  & 0--1\%   & 0--5\%     \\
\hline
\end{tabular}
\label{tab:syst}
\end{table}

Theoretical uncertainties affect all quantities derived from simulation such as the signal acceptance,
the transfer factors used in the estimate of the lost lepton and one-lepton backgrounds,
and the estimate of the \Znunu background.
The uncertainty resulting from missing higher-order corrections is estimated by varying the renormalization and factorization scales
by a factor of two~\cite{Catani:2003zt,Cacciari2003fi} with the two scales taken to be the same in each variation.
The effect of the uncertainties in the parton distribution functions is estimated using 100 variations provided with the NNPDF sets,
and the effect of the uncertainty in the value of the strong coupling constant is estimated by varying the value $\alpS(m_{\PZ})=0.1180$ by $\pm0.0015$~\cite{Butterworth:2015oua}.
All theory uncertainties are varied based on the NNPDF3.0 scheme.

The \ptmiss lineshape is corrected to account for mismodeling effects from \ptmiss resolution
and $\NISR$.
The uncertainty in these corrections results in a 1--50\% uncertainty in the estimated backgrounds, depending on signal region.
The uncertainty in the $\NISR$ rescaling also affects the signal acceptance.
The effect is small in most search regions, but can be noticeable in signal scenarios with a compressed mass spectrum.

The effect of the uncertainty in the jet energy scale is 1--34\% in the estimated backgrounds and up to 24\% in the signal acceptance.
Variations in the efficiency of the {\PQb} jet and soft {\PQb} object identification typically affect the estimated signal and background yields by 0.1\% and 3\%, with a full range up to 10\%.

The uncertainty in the cross section of $\wjets$ events with jets containing {\PQb} quarks is an important source of
uncertainty in the estimation of the $\wjets$ background.
A comparison of the multiplicity of {\PQb}-tagged jets between data and simulation is performed in a $\wjets$ enriched control sample obtained with the same selection as for the \Mlb validation test,
with the additional requirement of $\ptmiss>250\GeV$. From this study, we estimate
a 50\% uncertainty in the $\PW+\PQb(\cPaqb)$ cross section
resulting in a 20--40\% uncertainty in the $\wjets$ background estimate.

\section{Results and interpretation}
\label{sec:results}

\begin{figure*}[htb]
\centering
\includegraphics[width=0.98\textwidth]{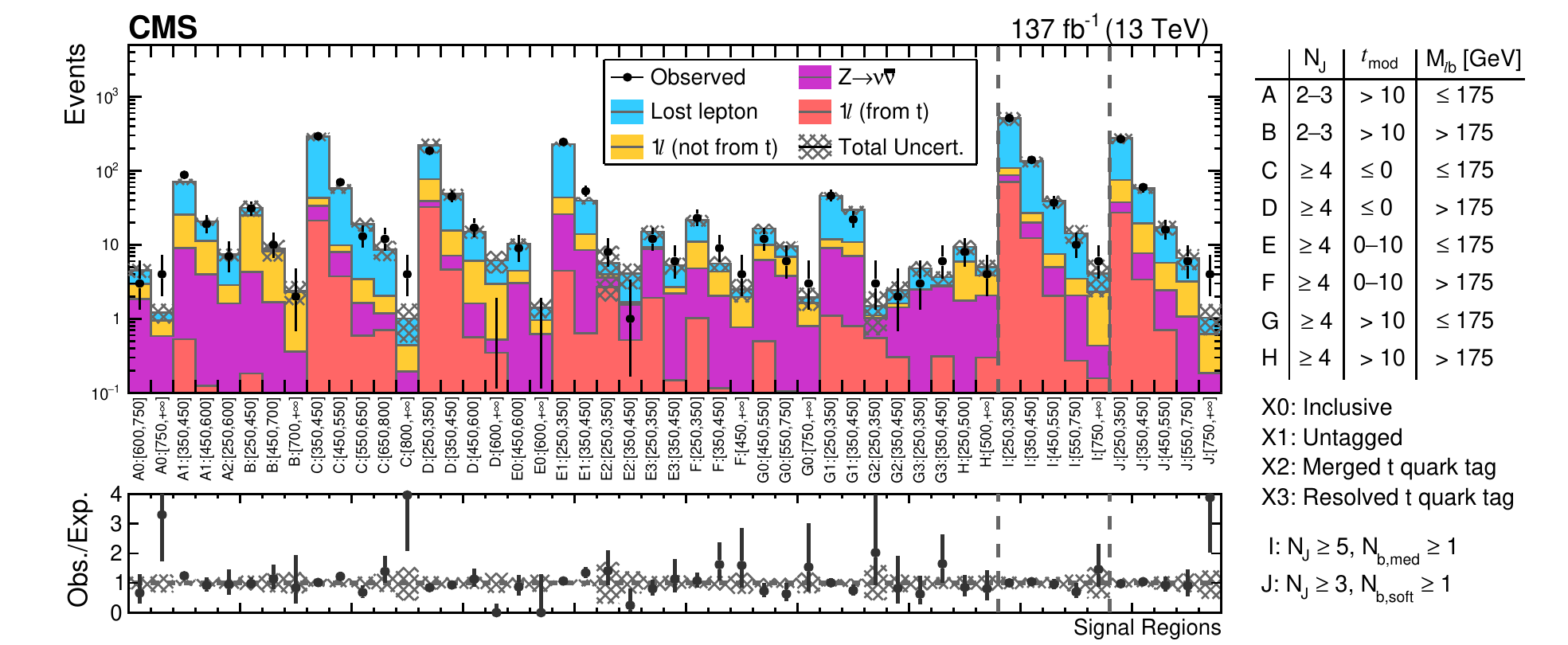}
\caption{
The observed and expected yields in Tables~\ref{tab:results-std} and~\ref{tab:results-dedi} and their ratios are shown as stacked histograms.
The lost lepton and 1$\ell$ (not from $\PQt$) are estimated from data-driven methods, while 1$\ell$ (from $\PQt$) and \Znunu backgrounds are taken from simulation.
The uncertainties consist of statistical and systematic components summed in quadrature and are shown as shaded bands.
}
\label{fig:results}
\end{figure*}

The event yields and the SM predictions in the search regions are summarized in Tables~\ref{tab:results-std} and~\ref{tab:results-dedi}.
These results are also illustrated in Fig.~\ref{fig:results}.
The observed yields are consistent with the estimated SM backgrounds.
Isolated fluctuations are observed in a few signal region bins.
The data events in these signal region bins were inspected carefully to determine if any detector or reconstruction effects were the source of the high \ptmiss. No such issues were detected.

\begin{table*}[htb]
\centering
\topcaption{
  The observed and expected yields in the standard search regions.
  For the top quark tagging categories, we use the abbreviations U for untagged, M for merged, and R for
  resolved.
}
\label{tab:results-std}
\cmsTable{
\begin{tabular}{ cccccccccccc }
\hline
\multirow{2}{*}{Label} & \multirow{2}{*}{$\NJ$}   & \multirow{2}{*}{$\tmod$}  & $\Mlb$                     & \PQt                & $\ptmiss$   & Lost           & 1$\ell$ (not    & 1$\ell$       & \multirow{2}{*}{\Znunu} & Total     & Total    \\
                       &                          &                           & [{\GeVns}]                 & cat.                & [{\GeVns}]      & lepton         & from $\PQt$)    & (from $\PQt$) &                                       & expected  & observed \\ \hline
\multirow{2}{*}{A0}    & \multirow{5}{*}{2--3}    & \multirow{5}{*}{$>$10   } & \multirow{5}{*}{$\le$175}  & \multirow{2}{*}{\NA} & 600--750       & $1.6 \pm 0.7$   & $1.1 \pm 0.5$   & $0.09\pm0.09$ & $1.8\pm0.4$   & $4.5 \pm 0.9$  & 3     \\
                       &                          &                           &                            &                     & 750--$+\infty$ & $0.26 \pm 0.19$ & $0.37 \pm 0.28$ &     \NA      & $0.59\pm0.20$ & $1.2 \pm 0.4$  & 4     \\
\multirow{2}{*}{A1}    &                          &                           &                            & \multirow{2}{*}{U}  & 350--450       & $46 \pm 5$      & $16 \pm 5$      & $0.5\pm0.5$   & $8.5\pm1.2$   & $71 \pm 8$     & 88    \\
                       &                          &                           &                            &                     & 450--600       & $9.4 \pm 1.5$   & $7.3 \pm 2.4$   & $0.12\pm0.12$ & $3.9\pm0.7$   & $20.7 \pm 3.0$ & 19    \\
\multirow{1}{*}{A2}    &                          &                           &                            & \multirow{1}{*}{M}  & 250--600       & $4.5 \pm 1.1$   & $1.2\pm0.4$     & $0.03\pm0.03$ & $1.6\pm0.4$   & $7.4 \pm 1.3$  & 7     \\ [\cmsTabSkip]
\multirow{3}{*}{B}     & \multirow{3}{*}{2--3}    & \multirow{3}{*}{$>$10   } & \multirow{3}{*}{$>$175}    & \multirow{3}{*}{\NA} & 250--450       & $6.6 \pm 1.5$   & $21 \pm 10$     & $0.18\pm0.18$ & $4.1\pm0.9$   & $32 \pm 11$    & 31    \\
                       &                          &                           &                            &                     & 450--700       & $0.55 \pm 0.26$ & $7 \pm 4$       &      \NA      & $1.7\pm0.5$   & $9 \pm 4$      & 10    \\
                       &                          &                           &                            &                     & 700--$+\infty$ & $0.07 \pm 0.06$ & $2.0 \pm 1.1$   &      \NA      & $0.36\pm0.15$ & $2.4 \pm 1.1$  & 2     \\ [\cmsTabSkip]
\multirow{5}{*}{C}     & \multirow{5}{*}{$\geq$4} & \multirow{5}{*}{$\le$0}   & \multirow{5}{*}{$\le$175}  & \multirow{5}{*}{\NA} & 350--450       & $245 \pm 23$    & $9.8 \pm 3.5$   & $21\pm21$     & $12.1\pm2.7$  & $289 \pm 32$   & 293   \\
                       &                          &                           &                            &                     & 450--550       & $48 \pm 7$      & $1.8 \pm 0.7$   & $4\pm4$       & $4.2\pm0.9$   & $58 \pm 8$     & 70    \\
                       &                          &                           &                            &                     & 550--650       & $16 \pm 4$      & $1.8 \pm 1.0$   & $0.6\pm0.6$   & $1.04\pm0.31$ & $19 \pm 4$     & 13    \\
                       &                          &                           &                            &                     & 650--800       & $6.6 \pm 2.5$   & $0.9 \pm 0.4$   & $0.7\pm0.7$   & $0.47\pm0.19$ & $8.6 \pm 2.6$  & 12    \\
                       &                          &                           &                            &                     & 800--$+\infty$ & $0.6 \pm 0.7$   & $0.25 \pm 0.13$ & $0.08\pm0.08$ & $0.12\pm0.08$ & $1.0 \pm 0.7$  & 4     \\ [\cmsTabSkip]
\multirow{4}{*}{D}     & \multirow{4}{*}{$\geq$4} & \multirow{4}{*}{$\le$0}   & \multirow{4}{*}{$>$175}    & \multirow{4}{*}{\NA} & 250--350       & $144 \pm 13$    & $38 \pm 13$     & $32\pm32$     & $6.5\pm1.5$   & $221 \pm 37$   & 186   \\
                       &                          &                           &                            &                     & 350--450       & $33 \pm 5$      & $8.3 \pm 3.4$   & $5\pm5$       & $2.5\pm0.7$   & $48 \pm 8$     & 45    \\
                       &                          &                           &                            &                     & 450--600       & $8.9 \pm 2.5$   & $4.5 \pm 1.9$   & $0.6\pm0.6$   & $1.05\pm0.26$ & $15.0 \pm 3.2$ & 17    \\
                       &                          &                           &                            &                     & 600--$+\infty$ & $3.2 \pm 2.1$   & $2.4 \pm 0.9$   & $0.35\pm0.35$ & $0.17\pm0.16$ & $6.2 \pm 2.4$  & 0     \\ [\cmsTabSkip]
\multirow{2}{*}{E0}    & \multirow{8}{*}{$\geq$4} & \multirow{8}{*}{0--10}    & \multirow{8}{*}{$\le$175}  & \multirow{2}{*}{\NA} & 450--600       & $5.9 \pm 1.5$   & $1.4 \pm 0.7$   &      \NA      & $3.0\pm0.7$   & $10.4 \pm 1.8$ & 9     \\
                       &                          &                           &                            &                     & 600--$+\infty$ & $0.45 \pm 0.28$ & $0.34 \pm 0.18$ &      \NA      & $0.62\pm0.24$ & $1.4 \pm 0.4$  & 0     \\
\multirow{2}{*}{E1}    &                          &                           &                            & \multirow{2}{*}{U}  & 250--350       & $186 \pm 17$    & $18 \pm 6$      & $4\pm4$       & $21\pm4$      & $230 \pm 19$   & 245   \\
                       &                          &                           &                            &                     & 350--450       & $26 \pm 4$      & $5.4 \pm 1.8$   & $0.6\pm0.6$   & $7.8\pm1.3$   & $40 \pm 4$     & 53    \\
\multirow{2}{*}{E2}    &                          &                           &                            & \multirow{2}{*}{M}  & 250--350       & $1.7 \pm 0.9$   & $0.38\pm0.16$   & $2.7\pm2.7$   & $0.95\pm0.27$ & $5.7 \pm 2.8$  & 8     \\
                       &                          &                           &                            &                     & 350--450       & $2.4 \pm 1.4$   & $0.12\pm0.12$   & $0.5\pm0.5$   & $1.05\pm0.29$ & $4.1 \pm 1.5$  & 1     \\
\multirow{2}{*}{E3}    &                          &                           &                            & \multirow{2}{*}{R}  & 250--350       & $5.6 \pm 1.8$   & $0.7\pm0.4$     & $1.9\pm1.9$   & $6.8\pm1.5$   & $15.0 \pm 3.0$ & 12    \\
                       &                          &                           &                            &                     & 350--450       & $2.6 \pm 1.4$   & $0.48\pm0.25$   & $0.15\pm0.15$ & $2.0\pm0.5$   & $5.3 \pm 1.5$  & 6     \\ [\cmsTabSkip]
\multirow{3}{*}{F}     & \multirow{3}{*}{$\geq$4} & \multirow{3}{*}{0--10}    & \multirow{3}{*}{$>$175}    & \multirow{3}{*}{\NA} & 250--350       & $10.4 \pm 2.5$  & $6.2 \pm 3.2$   & $1.0\pm1.0$   & $3.8\pm0.8$   & $21 \pm 4$     & 23    \\
                       &                          &                           &                            &                     & 350--450       & $1.2 \pm 0.9$   & $2.3 \pm 1.2$   & $0.12\pm0.12$ & $1.9\pm0.8$   & $5.6 \pm 1.7$  & 9     \\
                       &                          &                           &                            &                     & 450--$+\infty$ & $0.5^{+1.0}_{-0.5}$ & $1.2 \pm 0.7$ & $0.08\pm0.08$ & $0.69\pm0.25$ & $2.5 \pm 1.2$  & 4     \\ [\cmsTabSkip]
\multirow{3}{*}{G0}    & \multirow{9}{*}{$\geq$4} & \multirow{9}{*}{$>$10   } & \multirow{9}{*}{$\le$175}  & \multirow{3}{*}{\NA} & 450--550       & $6.5 \pm 1.9$   & $3.8 \pm 1.7$   & $0.5\pm0.5$   & $5.7\pm1.0$   & $16.6 \pm 2.8$ & 12    \\
                       &                          &                           &                            &                     & 550--750       & $2.7 \pm 1.2$   & $3.1 \pm 1.2$   & $0.1\pm0.1$   & $3.7\pm0.8$   & $9.5 \pm 1.9$  & 6     \\
                       &                          &                           &                            &                     & 750--$+\infty$ & $0.33 \pm 0.18$ & $0.83 \pm 0.35$ &      \NA      & $0.79\pm0.16$ & $1.9 \pm 0.4$  & 3     \\
\multirow{2}{*}{G1}    &                          &                           &                            & \multirow{2}{*}{U}  & 250--350       & $34 \pm 5$      & $2.8 \pm 1.2$   & $1.1\pm1.1$   & $7.9\pm1.8$   & $46 \pm 6$     & 46    \\
                       &                          &                           &                            &                     & 350--450       & $19 \pm 4$      & $3.8 \pm 1.6$   & $0.8\pm0.8$   & $6.3\pm1.5$   & $30 \pm 4$     & 22    \\
\multirow{2}{*}{G2}    &                          &                           &                            & \multirow{2}{*}{M}  & 250--350       & $0.37 \pm 0.27$ & $0.1\pm0.06$    & $0.6\pm0.6$   & $0.46\pm0.15$ & $1.5 \pm 0.6$  & 3     \\
                       &                          &                           &                            &                     & 350--450       & $0.8 \pm 0.5$   & $0.2\pm0.1$     & $0.3\pm0.3$   & $1.12\pm0.23$ & $2.4 \pm 0.6$  & 2     \\
\multirow{2}{*}{G3}    &                          &                           &                            & \multirow{2}{*}{R}  & 250--350       & $2.3 \pm 1.0$   & $0.06\pm0.09$   & $0.09\pm0.09$ & $2.4\pm0.5$   & $4.8 \pm 1.2$  & 3     \\
                       &                          &                           &                            &                     & 350--450       & $0.8 \pm 0.5$   & $0.12\pm0.08$   & $0.31\pm0.31$ & $2.4\pm0.6$   & $3.6 \pm 0.8$  & 6     \\ [\cmsTabSkip]
\multirow{2}{*}{H}     & \multirow{2}{*}{$\geq$4} & \multirow{2}{*}{$>$10   } & \multirow{2}{*}{$>$175}    & \multirow{2}{*}{\NA} & 250--500       & $3.4 \pm 1.4$   & $4.2 \pm 2.0$   & $0.09\pm0.09$ & $1.7\pm0.4$   & $9.4 \pm 2.5$  & 8     \\
                       &                          &                           &                            &                     & 500--$+\infty$ & $1.1 \pm 0.5$   & $1.8 \pm 1.0$   & $0.3\pm0.3$   & $1.8\pm0.6$   & $5.0 \pm 1.3$  & 4     \\
\hline
\end{tabular}}
\end{table*}

\begin{table*}[htb]
\centering
\topcaption{
The observed and expected yields for signal regions targeting scenarios of top squark production with a compressed mass spectrum.
}
\label{tab:results-dedi}
\cmsTable{
\begin{tabular}{ ccccccccccc }
\hline
\multirow{2}{*}{Label} & \multirow{2}{*}{$\NJ$}   & \multirow{2}{*}{$\Nmedb$} & \multirow{2}{*}{$\Nsoftb$} & $\ptmiss$  & Lost                 & 1$\ell$ (not           & 1$\ell$       & \multirow{2}{*}{\Znunu} & Total     & Total    \\
                       &                          &                           &                            & [{\GeVns}] & lepton               & from $\PQt$)           & (from $\PQt$) &                                               & expected  & observed \\ \hline
 \multirow{5}{*}{I}    & \multirow{5}{*}{$\geq$5} & \multirow{5}{*}{$\geq$1}  & \multirow{5}{*}{$\geq$0}   & 250--350       & $403 \pm 40$    & $21 \pm 8$         & $71\pm71$     & $17\pm4$      & $511 \pm 81$   & 513   \\
                       &                          &                           &                            & 350--450       & $108 \pm 15$    & $6.8 \pm 2.5$      & $12\pm12$     & $7.8\pm1.6$   & $134 \pm 19$   & 140   \\
                       &                          &                           &                            & 450--550       & $31 \pm 8$      & $2.5 \pm 1.0$      & $2.0\pm2.0$   & $2.9\pm0.8$   & $39 \pm 8$     & 37    \\
                       &                          &                           &                            & 550--750       & $11 \pm 5$      & $1.4 \pm 0.6$      & $0.27\pm0.27$ & $1.8\pm0.5$   & $14 \pm 5$     & 10    \\
                       &                          &                           &                            & 750--$+\infty$ & $1.8 \pm 1.1$   & $1.9^{+2.5}_{-1.9}$ & $0.16\pm0.16$ & $0.28\pm0.10$  & $4.1 \pm 2.5$  & 6    \\ [\cmsTabSkip]
 \multirow{5}{*}{J}    & \multirow{5}{*}{$\geq$3} & \multirow{5}{*}{$\geq$0}  & \multirow{5}{*}{$\geq$1}   & 250--350       & $201 \pm 21$    & $37 \pm 7$         & $27\pm27$     & $10.4\pm1.5$  & $276 \pm 35$   & 268   \\
                       &                          &                           &                            & 350--450       & $38 \pm 7$      & $11.6 \pm 2.2$     & $3.4\pm3.4$   & $4.3\pm0.9$   & $58 \pm 8$     & 60    \\
                       &                          &                           &                            & 450--550       & $11.5 \pm 3.5$  & $3.3 \pm 0.6$      & $0.7\pm0.7$   & $1.7\pm0.6$   & $17 \pm 4$     & 16    \\
                       &                          &                           &                            & 550--750       & $3.5 \pm 2.3$   & $2.1 \pm 0.5$      &      \NA      & $1.1\pm0.8$   & $6.6 \pm 2.5$  & 6     \\
                       &                          &                           &                            & 750--$+\infty$ & $0.4 \pm 0.4$   & $0.44 \pm 0.16$    & $0.02\pm0.02$ & $0.2\pm0.4$   & $1.0 \pm 0.6$  & 4     \\
\hline
\end{tabular}
}
\end{table*}

Results are interpreted in the context of top squark pair production models described in Section~\ref{sec:intro}.
For a given model, 95\% confidence level (CL) upper limits on the production cross sections are derived as a function of the mass of the SUSY particles. 
The search regions are combined using a modified frequentist approach, employing the $\CLs$ criterion and an asymptotic formulation~\cite{Junk:1999kv,Read:2002hq,Cowan:2010js,ATLAS:2011tau}.
The likelihood function is constructed by multiplying the probability density functions from each search region. 
These probability density functions are products of Poisson functions for the control region yields and log-normal constraint functions for the nuisance parameters, with correlated parameters among the search regions being accounted for.
When computing the limit, the expected signal yields are corrected for the possible contributions of signal events to the control samples.
These corrections are typically around 5--10\%.

For the models in which both top squarks decay to a top quark and an \lsp, the limits are derived from the
$\Delta m\left(\PSQt,\PSGczDo\right)\sim m_{\PW}$ search regions when $100\leq \Delta m\left(\PSQt,\PSGczDo\right) \leq 150\GeV$, and from the $\Delta m\left(\PSQt,\PSGczDo\right)\sim m_{\PQt}$ search regions when $150 \le \Delta m\left(\PSQt,\PSGczDo\right) \le 225\GeV$. For all other models, the cross section limits are obtained from the standard search regions.

In the case of $\Delta m\left(\PSQt,\PSGczDo\right)\sim m_{\PW}$, the specially designed signal
regions result in improvements of up to a factor of five in cross section sensitivity
with respect to the results that would have been obtained based on the standard search regions.
On the other hand,
the corresponding improvements from the signal regions designed for
$\Delta m\left(\PSQt,\PSGczDo\right)\sim m_{\PQt}$ are typically of the order of
10--20\%. In the high mass region, this analysis is sensitive to an additional $\sim$200\GeV in expected limit for top squark masses~\cite{Sirunyan:2017xse}.

\begin{figure*}[htb]
\centering
\includegraphics[width=0.95\textwidth]{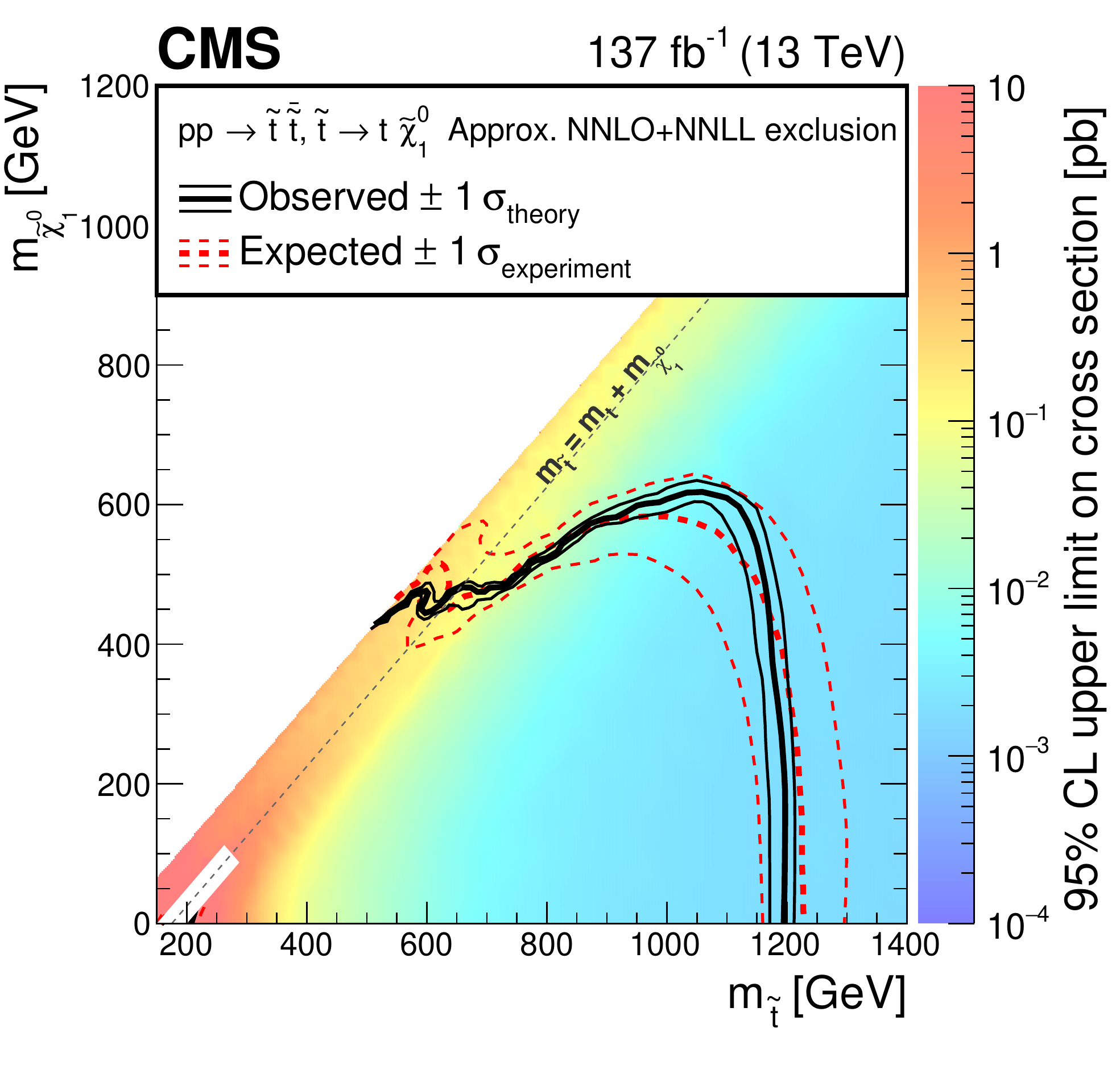}
\caption{
Exclusion limits at 95\%~\CL for the $\Pp\Pp\to\PSQt\PASQt\to \PQt\PAQt \PSGczDo\PSGczDo$ scenario.
The colored map illustrates the 95\% CL upper limits on the product of the production cross section and branching fraction.
The area enclosed by the thick black curve represents the observed exclusion region, and that enclosed by the thick, dashed red curve represents the expected exclusion.
The thin dotted (red) curves indicate the region containing 68\% of the distribution of limits expected under the background-only hypothesis. The thin solid (black) curves show the change in the observed limit by varying the signal cross sections within their theoretical uncertainties.
The white band excluded from the limits corresponds to the region $\abs{m_{\PSQt}-m_{\mathrm{t}}-m_{\PSGczDo}} < 25\GeV,\, m_{\PSQt}<275\GeV$, where the selection acceptance for top squark pair production changes rapidly and is therefore very sensitive to the details of the simulation.
}
\label{fig:limits:T2tt}
\end{figure*}

The 95\% \CL upper limits on cross sections for the $\Pp\Pp\to\PSQt\PASQt\to \PQt\PAQt\PSGczDo\PSGczDo$ process, as a function of sparticle
masses and assuming that the top quarks are not
polarized, are shown in Fig.~\ref{fig:limits:T2tt}.
In this figure we also
show the excluded region of parameter space based on the expected
cross section for top squark pair production.
We exclude the existence of top squarks with masses up to 1.2\TeV for a massless neutralino,
and neutralinos with masses up to 600\GeV for $m_{\PSQt}=1\TeV$.
The most sensitive search regions for these processes are those with high \tmod and low \Mlb values.
Signal models with higher $\Delta m\left(\PSQt,\PSGczDo\right)$ are more sensitive in the regions with higher \ptmiss.
The white band corresponds to the region $\abs{m_{\PSQt}-m_{\mathrm{t}}-m_{\PSGczDo}} < 25\GeV,\, m_{\PSQt}<275\GeV$, where the selection acceptance for top squark pair production changes rapidly.
In this region the acceptance is very sensitive to the details of the simulation, and therefore no interpretation is performed.

\begin{figure*}[htb]
\centering
\includegraphics[width=0.95\textwidth]{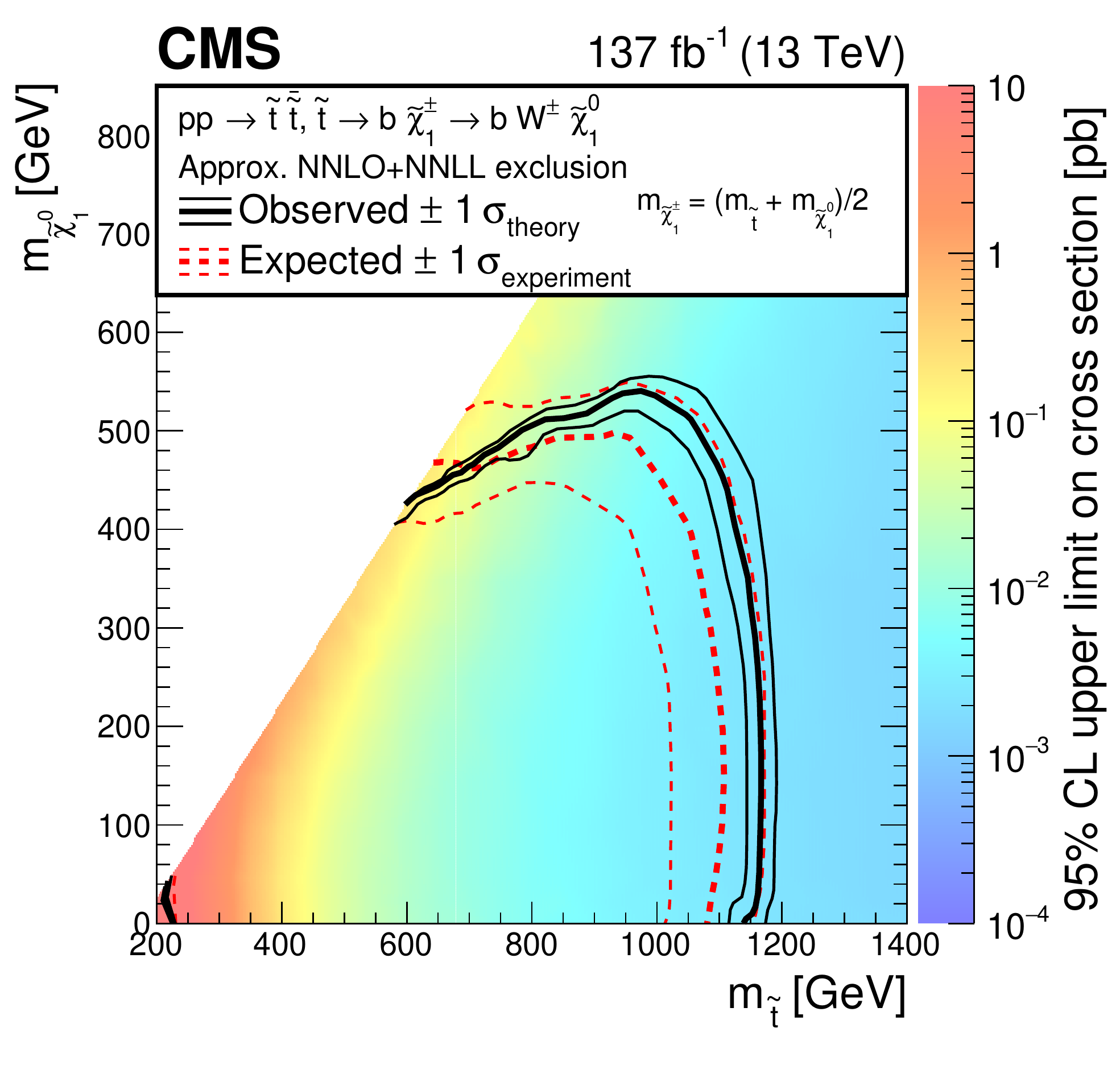}
\caption{
Exclusion limits at 95\%~\CL for the $\Pp\Pp\to\PSQt\PASQt\to \PQb\PAQb\PSGcpmDo\PSGcpmDo \left(\PSGcpmDo\to\PW\PSGczDo\right)$ scenario.
The mass of \chgo is chosen to be $(m_{\PSQt} + m_{\PSGczDo})/2$.
The colored map illustrates the 95\% CL upper limits on the product of the production cross section and branching fraction.
The area enclosed by the thick black curve represents the observed exclusion region, and that enclosed by the thick, dashed red curve represents the expected exclusion.
The thin dotted (red) curves indicate the region containing 68\% of the distribution of limits expected under the background-only hypothesis. The thin solid (black) curves show the change in the observed limit by varying the signal cross sections within their theoretical uncertainties.
}
\label{fig:limits:T2bW}
\end{figure*}

\begin{figure*}[htb]
\centering
\includegraphics[width=0.95\textwidth]{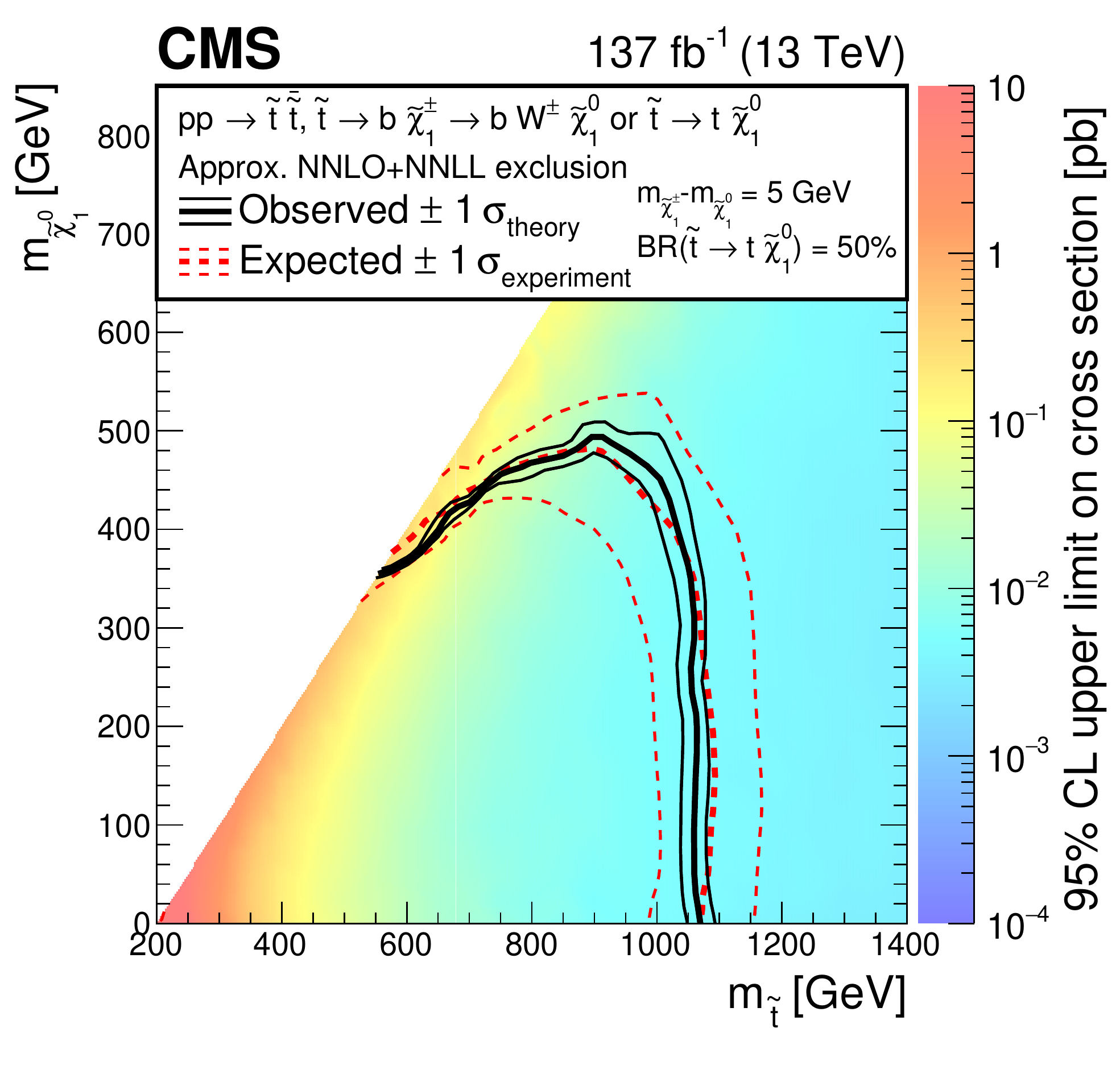}
\caption{
Exclusion limits at 95\%~\CL for the $\Pp\Pp\to\PSQt\PASQt\to\PQt{\PQb}\PSGcpmDo\PSGczDo \left(\PSGcpmDo\to\PW^{*}\PSGczDo\right)$ scenario.
The mass difference between the \chgo and the \PSGczDo is taken to be 5\GeV.
The colored map illustrates the 95\% CL upper limits on the product of the production cross section and branching fraction.
The area enclosed by the thick black curve represents the observed exclusion region, and that enclosed by the thick, dashed red curve represents the expected exclusion.
The thin dotted (red) curves indicate the region containing 68\% of the distribution of limits expected under the background-only hypothesis. The thin solid (black) curves show the change in the observed limit by varying the signal cross sections within their theoretical uncertainties.
}
\label{fig:limits:T2bt}
\end{figure*}

Figures~\ref{fig:limits:T2bW} and \ref{fig:limits:T2bt} display the equivalent limits
for the $\Pp\Pp\to\PSQt\PASQt \to \PQb\PAQb\PSGcpmDo\PSGcpmDo \left(\PSGcpmDo\to\PW\PSGczDo\right)$ and
$\Pp\Pp\to\PSQt\PASQt \to \PQt\PQb\PSGcpmDo\PSGczDo  \left(\PSGcpmDo\to\PW^{*}\PSGczDo\right)$ scenarios, respectively.
The search regions with high \Mlb are most sensitive to these models.
These models are characterized by three mass parameters (for the
top squark, the chargino, and the neutralino).
In the mixed decay scenario
of Fig.~\ref{fig:limits:T2bt},
we have assumed a compressed mass spectrum
for the neutralino-chargino pair, which is theoretically
favored if the $\PSGcpmDo$ and the $\PSGczDo$ are higgsinos.
The search has very poor sensitivity for models with
this mass spectrum when both top squarks decay to charginos.
Therefore in the case of Fig.~\ref{fig:limits:T2bW}, we have chosen
a larger mass splitting between the $\PSGcpmDo$ and the $\PSGczDo$.

\section{Summary}
\label{sec:summary}
A search for direct top squark pair production is performed using events with one lepton, jets, and significant missing transverse momentum.
The search is based on proton-proton collision data at a center-of-mass energy of 13\TeV recorded by the CMS experiment at the LHC during 2016-2018 and corresponding to an integrated luminosity of \Lint.
The leading backgrounds in this analysis, mainly dileptonic \ttbar decays, where one of the leptons is not reconstructed or identified, and $\wjets$ production are estimated from data control regions. The semileptonic $\ttbar$ and $\Znunu$ backgrounds are taken from simulation.
No significant deviations from the standard model expectations are observed. Limits on pair-produced top squarks are established in the context of supersymmetry models conserving $R$-parity.
Exclusion limits at 95\%~\CL for top squark masses up to 1.2\TeV are set for a massless neutralino. For models with a top squark mass of 1\TeV, neutralino masses up to 600\GeV are excluded.

\clearpage
\begin{acknowledgments}
We congratulate our colleagues in the CERN accelerator departments for the excellent performance of the LHC and thank the technical and administrative staffs at CERN and at other CMS institutes for their contributions to the success of the CMS effort. In addition, we gratefully acknowledge the computing centers and personnel of the Worldwide LHC Computing Grid for delivering so effectively the computing infrastructure essential to our analyses. Finally, we acknowledge the enduring support for the construction and operation of the LHC and the CMS detector provided by the following funding agencies: BMBWF and FWF (Austria); FNRS and FWO (Belgium); CNPq, CAPES, FAPERJ, FAPERGS, and FAPESP (Brazil); MES (Bulgaria); CERN; CAS, MoST, and NSFC (China); COLCIENCIAS (Colombia); MSES and CSF (Croatia); RPF (Cyprus); SENESCYT (Ecuador); MoER, ERC IUT, PUT and ERDF (Estonia); Academy of Finland, MEC, and HIP (Finland); CEA and CNRS/IN2P3 (France); BMBF, DFG, and HGF (Germany); GSRT (Greece); NKFIA (Hungary); DAE and DST (India); IPM (Iran); SFI (Ireland); INFN (Italy); MSIP and NRF (Republic of Korea); MES (Latvia); LAS (Lithuania); MOE and UM (Malaysia); BUAP, CINVESTAV, CONACYT, LNS, SEP, and UASLP-FAI (Mexico); MOS (Montenegro); MBIE (New Zealand); PAEC (Pakistan); MSHE and NSC (Poland); FCT (Portugal); JINR (Dubna); MON, RosAtom, RAS, RFBR, and NRC KI (Russia); MESTD (Serbia); SEIDI, CPAN, PCTI, and FEDER (Spain); MOSTR (Sri Lanka); Swiss Funding Agencies (Switzerland); MST (Taipei); ThEPCenter, IPST, STAR, and NSTDA (Thailand); TUBITAK and TAEK (Turkey); NASU (Ukraine); STFC (United Kingdom); DOE and NSF (USA).

\hyphenation{Rachada-pisek} Individuals have received support from the Marie-Curie program and the European Research Council and Horizon 2020 Grant, contract Nos.\ 675440, 752730, and 765710 (European Union); the Leventis Foundation; the A.P.\ Sloan Foundation; the Alexander von Humboldt Foundation; the Belgian Federal Science Policy Office; the Fonds pour la Formation \`a la Recherche dans l'Industrie et dans l'Agriculture (FRIA-Belgium); the Agentschap voor Innovatie door Wetenschap en Technologie (IWT-Belgium); the F.R.S.-FNRS and FWO (Belgium) under the ``Excellence of Science -- EOS'' -- be.h project n.\ 30820817; the Beijing Municipal Science \& Technology Commission, No. Z181100004218003; the Ministry of Education, Youth and Sports (MEYS) of the Czech Republic; the Lend\''ulet (``Momentum'') Program and the J\'anos Bolyai Research Scholarship of the Hungarian Academy of Sciences, the New National Excellence Program \'UNKP, the NKFIA research grants 123842, 123959, 124845, 124850, 125105, 128713, 128786, and 129058 (Hungary); the Council of Science and Industrial Research, India; the HOMING PLUS program of the Foundation for Polish Science, cofinanced from European Union, Regional Development Fund, the Mobility Plus program of the Ministry of Science and Higher Education, the National Science Center (Poland), contracts Harmonia 2014/14/M/ST2/00428, Opus 2014/13/B/ST2/02543, 2014/15/B/ST2/03998, and 2015/19/B/ST2/02861, Sonata-bis 2012/07/E/ST2/01406; the National Priorities Research Program by Qatar National Research Fund; the Ministry of Science and Education, grant no. 3.2989.2017 (Russia); the Programa Estatal de Fomento de la Investigaci{\'o}n Cient{\'i}fica y T{\'e}cnica de Excelencia Mar\'{\i}a de Maeztu, grant MDM-2015-0509 and the Programa Severo Ochoa del Principado de Asturias; the Thalis and Aristeia programs cofinanced by EU-ESF and the Greek NSRF; the Rachadapisek Sompot Fund for Postdoctoral Fellowship, Chulalongkorn University and the Chulalongkorn Academic into Its 2nd Century Project Advancement Project (Thailand); the Welch Foundation, contract C-1845; and the Weston Havens Foundation (USA).
\end{acknowledgments}

\bibliography{auto_generated}

\providecommand{\href}[2]{#2}\begingroup\raggedright\begin{thebibliography}{10}%
\makeatletter
\providecommand{\hrefCMSnoop }[0]{\@secondoftwo}%
\makeatother
\providecommand{\doi}{\texttt{doi:}\begingroup \urlstyle{tt}\Url}

\bibitem{Ramond:1971gb}
\hrefCMSnoop {}{P.~Ramond, ``Dual theory for free fermions'',} \textit{ Phys.
  Rev. D} \textbf{ 3} (1971) 2415,
\href{http://dx.doi.org/10.1103/PhysRevD.3.2415}{\doi{10.1103/PhysRevD.3.2415}}.

\bibitem{Golfand:1971iw}
\hrefCMSnoop {}{Y.~A. Gol'fand and E.~P. Likhtman, ``Extension of the algebra
  of {Poincare} group generators and violation of {P} invariance'',} \textit{
  JETP Lett.} \textbf{ 13} (1971) 323.
[Pisma Zh. Eksp. Teor. Fiz. 13 (1971) 452].

\bibitem{Neveu:1971rx}
\hrefCMSnoop {}{A.~Neveu and J.~H. Schwarz, ``Factorizable dual model of
  pions'',} \textit{ Nucl. Phys. B} \textbf{ 31} (1971) 86,
\href{http://dx.doi.org/10.1016/0550-3213(71)90448-2}{\doi{10.1016/0550-3213(71)90448-2}}.

\bibitem{Volkov:1972jx}
\hrefCMSnoop {}{D.~V. Volkov and V.~P. Akulov, ``{Possible universal neutrino
  interaction}'',} \textit{ JETP Lett.} \textbf{ 16} (1972) 438.
[Pisma Zh. Eksp. Teor. Fiz. 16 (1972) 621].

\bibitem{Wess:1973kz}
\hrefCMSnoop {}{J.~Wess and B.~Zumino, ``A {Lagrangian} model invariant under
  supergauge transformations'',} \textit{ Phys. Lett. B} \textbf{ 49} (1974)
  52,
\href{http://dx.doi.org/10.1016/0370-2693(74)90578-4}{\doi{10.1016/0370-2693(74)90578-4}}.

\bibitem{Wess:1974tw}
\hrefCMSnoop {}{J.~Wess and B.~Zumino, ``Supergauge transformations in
  four-dimensions'',} \textit{ Nucl. Phys. B} \textbf{ 70} (1974) 39,
\href{http://dx.doi.org/10.1016/0550-3213(74)90355-1}{\doi{10.1016/0550-3213(74)90355-1}}.

\bibitem{Fayet:1974pd}
\hrefCMSnoop {}{P.~Fayet, ``Supergauge invariant extension of the {Higgs}
  mechanism and a model for the electron and its neutrino'',} \textit{ Nucl.
  Phys. B} \textbf{ 90} (1975) 104,
\href{http://dx.doi.org/10.1016/0550-3213(75)90636-7}{\doi{10.1016/0550-3213(75)90636-7}}.

\bibitem{Nilles:1983ge}
\hrefCMSnoop {}{H.~P. Nilles, ``Supersymmetry, supergravity and particle
  physics'',} \textit{ Phys. Rept.} \textbf{ 110} (1984) 1,
\href{http://dx.doi.org/10.1016/0370-1573(84)90008-5}{\doi{10.1016/0370-1573(84)90008-5}}.

\bibitem{Farrar:1978xj}
\hrefCMSnoop {}{G.~R. Farrar and P.~Fayet, ``Phenomenology of the production,
  decay, and detection of new hadronic states associated with supersymmetry'',}
  \textit{ Phys. Lett. B} \textbf{ 76} (1978) 575,
\href{http://dx.doi.org/10.1016/0370-2693(78)90858-4}{\doi{10.1016/0370-2693(78)90858-4}}.

\bibitem{Jungman:1995df}
\hrefCMSnoop {}{G.~Jungman, M.~Kamionkowski, and K.~Griest, ``Supersymmetric
  dark matter'',} \textit{ Phys. Rept.} \textbf{ 267} (1996) 195,
  \href{http://dx.doi.org/10.1016/0370-1573(95)00058-5}{\doi{10.1016/0370-1573(95)00058-5}},
\href{http://www.arXiv.org/abs/hep-ph/9506380}{\texttt{arXiv:hep-ph/9506380}}.

\bibitem{tHooft:1979rat}
\hrefCMSnoop {}{G.~'t~Hooft, ``Naturalness, chiral symmetry, and spontaneous
  chiral symmetry breaking'',} \textit{ NATO Sci. Ser. B} \textbf{ 59} (1980)
  135,
\href{http://dx.doi.org/10.1007/978-1-4684-7571-5_9}{\doi{10.1007/978-1-4684-7571-5_9}}.

\bibitem{Dine:1981za}
\hrefCMSnoop {}{M.~Dine, W.~Fischler, and M.~Srednicki, ``Supersymmetric
  technicolor'',} \textit{ Nucl. Phys. B} \textbf{ 189} (1981) 575,
\href{http://dx.doi.org/10.1016/0550-3213(81)90582-4}{\doi{10.1016/0550-3213(81)90582-4}}.

\bibitem{Dimopoulos:1981au}
\hrefCMSnoop {}{S.~Dimopoulos and S.~Raby, ``Supercolor'',} \textit{ Nucl.
  Phys. B} \textbf{ 192} (1981) 353,
\href{http://dx.doi.org/10.1016/0550-3213(81)90430-2}{\doi{10.1016/0550-3213(81)90430-2}}.

\bibitem{Dimopoulos:1981zb}
\hrefCMSnoop {}{S.~Dimopoulos and H.~Georgi, ``Softly broken supersymmetry and
  {SU(5)}'',} \textit{ Nucl. Phys. B} \textbf{ 193} (1981) 150,
\href{http://dx.doi.org/10.1016/0550-3213(81)90522-8}{\doi{10.1016/0550-3213(81)90522-8}}.

\bibitem{Kaul:1981hi}
\hrefCMSnoop {}{R.~K. Kaul and P.~Majumdar, ``Cancellation of quadratically
  divergent mass corrections in globally supersymmetric spontaneously broken
  gauge theories'',} \textit{ Nucl. Phys. B} \textbf{ 199} (1982) 36,
\href{http://dx.doi.org/10.1016/0550-3213(82)90565-X}{\doi{10.1016/0550-3213(82)90565-X}}.

\bibitem{Aaboud:2016lwz}
\hrefCMSnoop {}{{ATLAS Collaboration}, ``Search for top squarks in final states
  with one isolated lepton, jets, and missing transverse momentum in
  {$\sqrt{s}=13\TeV$} {$\Pp\Pp$} collisions with the {ATLAS} detector'',}
  \textit{ Phys. Rev. D} \textbf{ 94} (2016) 052009,
  \href{http://dx.doi.org/10.1103/PhysRevD.94.052009}{\doi{10.1103/PhysRevD.94.052009}},
\href{http://www.arXiv.org/abs/1606.03903}{\texttt{arXiv:1606.03903}}.

\bibitem{Aaboud:2016uth}
\hrefCMSnoop {}{{ATLAS Collaboration}, ``Search for heavy long-lived charged
  {R-hadrons} with the {ATLAS} detector in {${3.2\fbinv}$} of proton--proton
  collision data at {$\sqrt{s} = 13\TeV$}'',} \textit{ Phys. Lett. B} \textbf{
  760} (2016) 647,
  \href{http://dx.doi.org/10.1016/j.physletb.2016.07.042}{\doi{10.1016/j.physletb.2016.07.042}},
\href{http://www.arXiv.org/abs/1606.05129}{\texttt{arXiv:1606.05129}}.

\bibitem{Aaboud:2017ejf}
\hrefCMSnoop {}{{ATLAS Collaboration}, ``Search for direct top squark pair
  production in events with a {Higgs} or {$\PZ$} boson, and missing transverse
  momentum in {$\sqrt{s}=13\TeV$} {$\Pp\Pp$} collisions with the {ATLAS}
  detector'',} \textit{ JHEP} \textbf{ 08} (2017) 006,
  \href{http://dx.doi.org/10.1007/JHEP08(2017)006}{\doi{10.1007/JHEP08(2017)006}},
\href{http://www.arXiv.org/abs/1706.03986}{\texttt{arXiv:1706.03986}}.

\bibitem{Aaboud:2017nfd}
\hrefCMSnoop {}{{ATLAS Collaboration}, ``Search for direct top squark pair
  production in final states with two leptons in {$\sqrt{s} = 13\TeV$}
  {$\Pp\Pp$} collisions with the {ATLAS} detector'',} \textit{ Eur. Phys. J. C}
  \textbf{ 77} (2017) 898,
  \href{http://dx.doi.org/10.1140/epjc/s10052-017-5445-x}{\doi{10.1140/epjc/s10052-017-5445-x}},
\href{http://www.arXiv.org/abs/1708.03247}{\texttt{arXiv:1708.03247}}.

\bibitem{Aaboud:2017ayj}
\hrefCMSnoop {}{{ATLAS Collaboration}, ``Search for a scalar partner of the top
  quark in the jets plus missing transverse momentum final state at
  {$\sqrt{s}=13\TeV$} with the {ATLAS} detector'',} \textit{ JHEP} \textbf{ 12}
  (2017) 085,
  \href{http://dx.doi.org/10.1007/JHEP12(2017)085}{\doi{10.1007/JHEP12(2017)085}},
\href{http://www.arXiv.org/abs/1709.04183}{\texttt{arXiv:1709.04183}}.

\bibitem{Aaboud:2017opj}
\hrefCMSnoop {}{{ATLAS Collaboration}, ``Search for {$B-L$}
  {R-parity-violating} top squarks in {$\sqrt{s} =13\TeV$} {$\Pp\Pp$}
  collisions with the {ATLAS} experiment'',} \textit{ Phys. Rev. D} \textbf{
  97} (2018) 032003,
  \href{http://dx.doi.org/10.1103/PhysRevD.97.032003}{\doi{10.1103/PhysRevD.97.032003}},
\href{http://www.arXiv.org/abs/1710.05544}{\texttt{arXiv:1710.05544}}.

\bibitem{Aaboud:2017nmi}
\hrefCMSnoop {}{{ATLAS Collaboration}, ``A search for pair-produced resonances
  in four-jet final states at {$\sqrt{s} = 13\TeV$} with the {ATLAS}
  detector'',} \textit{ Eur. Phys. J. C} \textbf{ 78} (2018) 250,
  \href{http://dx.doi.org/10.1140/epjc/s10052-018-5693-4}{\doi{10.1140/epjc/s10052-018-5693-4}},
\href{http://www.arXiv.org/abs/1710.07171}{\texttt{arXiv:1710.07171}}.

\bibitem{Aaboud:2017aeu}
\hrefCMSnoop {}{{ATLAS Collaboration}, ``Search for top-squark pair production
  in final states with one lepton, jets, and missing transverse momentum using
  {$36\fbinv$} of {$\sqrt{s}=13\TeV$} {$\Pp\Pp$} collision data with the
  {ATLAS} detector'',} \textit{ JHEP} \textbf{ 06} (2018) 108,
  \href{http://dx.doi.org/10.1007/JHEP06(2018)108}{\doi{10.1007/JHEP06(2018)108}},
\href{http://www.arXiv.org/abs/1711.11520}{\texttt{arXiv:1711.11520}}.

\bibitem{Aaboud:2018kya}
\hrefCMSnoop {}{{ATLAS Collaboration}, ``Search for top squarks decaying to tau
  sleptons in {$\Pp\Pp$} collisions at {$\sqrt{s}= 13\TeV$} with the {ATLAS}
  detector'',} \textit{ Phys. Rev. D} \textbf{ 98} (2018) 032008,
  \href{http://dx.doi.org/10.1103/PhysRevD.98.032008}{\doi{10.1103/PhysRevD.98.032008}},
\href{http://www.arXiv.org/abs/1803.10178}{\texttt{arXiv:1803.10178}}.

\bibitem{Aaboud:2018zjf}
\hrefCMSnoop {}{{ATLAS Collaboration}, ``Search for supersymmetry in final
  states with charm jets and missing transverse momentum in {$13\TeV$}
  {$\Pp\Pp$} collisions with the {ATLAS} detector'',} \textit{ JHEP} \textbf{
  09} (2018) 050,
  \href{http://dx.doi.org/10.1007/JHEP09(2018)050}{\doi{10.1007/JHEP09(2018)050}},
\href{http://www.arXiv.org/abs/1805.01649}{\texttt{arXiv:1805.01649}}.

\bibitem{Khachatryan:2016pxa}
\hrefCMSnoop {}{{CMS Collaboration}, ``Search for top squark pair production in
  compressed-mass-spectrum scenarios in proton-proton collisions at {$\sqrt{s}
  = 8\TeV$} using the {$\alpha_{\mathrm{T}}$} variable'',} \textit{ Phys. Lett.
  B} \textbf{ 767} (2017) 403,
  \href{http://dx.doi.org/10.1016/j.physletb.2017.02.007}{\doi{10.1016/j.physletb.2017.02.007}},
\href{http://www.arXiv.org/abs/1605.08993}{\texttt{arXiv:1605.08993}}.

\bibitem{Sirunyan:2016jpr}
\hrefCMSnoop {}{{CMS Collaboration}, ``Searches for pair production of
  third-generation squarks in {$\sqrt{s}=13\TeV$} {$\Pp\Pp$} collisions'',}
  \textit{ Eur. Phys. J. C} \textbf{ 77} (2017) 327,
  \href{http://dx.doi.org/10.1140/epjc/s10052-017-4853-2}{\doi{10.1140/epjc/s10052-017-4853-2}},
\href{http://www.arXiv.org/abs/1612.03877}{\texttt{arXiv:1612.03877}}.

\bibitem{Khachatryan:2017rhw}
\hrefCMSnoop {}{{CMS Collaboration}, ``Search for supersymmetry in the
  all-hadronic final state using top quark tagging in {$\Pp\Pp$} collisions at
  {$\sqrt{s} = 13\TeV$}'',} \textit{ Phys. Rev. D} \textbf{ 96} (2017) 012004,
  \href{http://dx.doi.org/10.1103/PhysRevD.96.012004}{\doi{10.1103/PhysRevD.96.012004}},
\href{http://www.arXiv.org/abs/1701.01954}{\texttt{arXiv:1701.01954}}.

\bibitem{Sirunyan:2017xse}
\hrefCMSnoop {}{{CMS Collaboration}, ``Search for top squark pair production in
  pp collisions at {$\sqrt{s}=13\TeV$} using single lepton events'',} \textit{
  JHEP} \textbf{ 10} (2017) 019,
  \href{http://dx.doi.org/10.1007/JHEP10(2017)019}{\doi{10.1007/JHEP10(2017)019}},
\href{http://www.arXiv.org/abs/1706.04402}{\texttt{arXiv:1706.04402}}.

\bibitem{Sirunyan:2017wif}
\hrefCMSnoop {}{{CMS Collaboration}, ``Search for direct production of
  supersymmetric partners of the top quark in the all-jets final state in
  proton-proton collisions at {$\sqrt{s}=13\TeV$}'',} \textit{ JHEP} \textbf{
  10} (2017) 005,
  \href{http://dx.doi.org/10.1007/JHEP10(2017)005}{\doi{10.1007/JHEP10(2017)005}},
\href{http://www.arXiv.org/abs/1707.03316}{\texttt{arXiv:1707.03316}}.

\bibitem{Sirunyan:2017kiw}
\hrefCMSnoop {}{{CMS Collaboration}, ``Search for the pair production of
  third-generation squarks with two-body decays to a bottom or charm quark and
  a neutralino in proton-proton collisions at {$\sqrt{s} = 13\TeV$}'',}
  \textit{ Phys. Lett. B} \textbf{ 778} (2018) 263,
  \href{http://dx.doi.org/10.1016/j.physletb.2018.01.012}{\doi{10.1016/j.physletb.2018.01.012}},
\href{http://www.arXiv.org/abs/1707.07274}{\texttt{arXiv:1707.07274}}.

\bibitem{Sirunyan:2017pjw}
\hrefCMSnoop {}{{CMS Collaboration}, ``Search for supersymmetry in
  proton-proton collisions at {$13\TeV$} using identified top quarks'',}
  \textit{ Phys. Rev. D} \textbf{ 97} (2018) 012007,
  \href{http://dx.doi.org/10.1103/PhysRevD.97.012007}{\doi{10.1103/PhysRevD.97.012007}},
\href{http://www.arXiv.org/abs/1710.11188}{\texttt{arXiv:1710.11188}}.

\bibitem{Sirunyan:2017leh}
\hrefCMSnoop {}{{CMS Collaboration}, ``Search for top squarks and dark matter
  particles in opposite-charge dilepton final states at {$\sqrt{s}=
  13\TeV$}'',} \textit{ Phys. Rev. D} \textbf{ 97} (2018) 032009,
  \href{http://dx.doi.org/10.1103/PhysRevD.97.032009}{\doi{10.1103/PhysRevD.97.032009}},
\href{http://www.arXiv.org/abs/1711.00752}{\texttt{arXiv:1711.00752}}.

\bibitem{Sirunyan:2018iwl}
\hrefCMSnoop {}{{CMS Collaboration}, ``Search for new physics in events with
  two soft oppositely charged leptons and missing transverse momentum in
  proton-proton collisions at {$\sqrt{s}= 13\TeV$}'',} \textit{ Phys. Lett. B}
  \textbf{ 782} (2018) 440,
  \href{http://dx.doi.org/10.1016/j.physletb.2018.05.062}{\doi{10.1016/j.physletb.2018.05.062}},
\href{http://www.arXiv.org/abs/1801.01846}{\texttt{arXiv:1801.01846}}.

\bibitem{Sirunyan:2018omt}
\hrefCMSnoop {}{{CMS Collaboration}, ``Search for top squarks decaying via
  four-body or chargino-mediated modes in single-lepton final states in
  proton-proton collisions at {$\sqrt{s} = 13\TeV$}'',} \textit{ JHEP} \textbf{
  09} (2018) 065,
  \href{http://dx.doi.org/10.1007/JHEP09(2018)065}{\doi{10.1007/JHEP09(2018)065}},
\href{http://www.arXiv.org/abs/1805.05784}{\texttt{arXiv:1805.05784}}.

\bibitem{Sirunyan:2018rlj}
\hrefCMSnoop {}{{CMS Collaboration}, ``Search for pair-produced resonances
  decaying to quark pairs in proton-proton collisions at {$\sqrt{s}=
  13\TeV$}'',} \textit{ Phys. Rev. D} \textbf{ 98} (2018) 112014,
  \href{http://dx.doi.org/10.1103/PhysRevD.98.112014}{\doi{10.1103/PhysRevD.98.112014}},
\href{http://www.arXiv.org/abs/1808.03124}{\texttt{arXiv:1808.03124}}.

\bibitem{Sirunyan:2018ell}
\hrefCMSnoop {}{{CMS Collaboration}, ``Inclusive search for supersymmetry in pp
  collisions at {$ \sqrt{s}=13\TeV$} using razor variables and boosted object
  identification in zero and one lepton final states'',} \textit{ JHEP}
  \textbf{ 03} (2019) 031,
  \href{http://dx.doi.org/10.1007/JHEP03(2019)031}{\doi{10.1007/JHEP03(2019)031}},
\href{http://www.arXiv.org/abs/1812.06302}{\texttt{arXiv:1812.06302}}.

\bibitem{Sirunyan:2019zyu}
\hrefCMSnoop {}{{CMS Collaboration}, ``Search for the pair production of light
  top squarks in the {$\Pepm\PGm^{\mp}$} final state in proton-proton
  collisions at {$\sqrt{s} = 13\TeV$}'',} \textit{ JHEP} \textbf{ 03} (2019)
  101,
  \href{http://dx.doi.org/10.1007/JHEP03(2019)101}{\doi{10.1007/JHEP03(2019)101}},
\href{http://www.arXiv.org/abs/1901.01288}{\texttt{arXiv:1901.01288}}.

\bibitem{Chatrchyan:2008zzk}
\hrefCMSnoop {}{{CMS Collaboration}, ``The {CMS} experiment at the {CERN}
  {LHC}'',} \textit{ JINST} \textbf{ 3} (2008) S08004,
\href{http://dx.doi.org/10.1088/1748-0221/3/08/S08004}{\doi{10.1088/1748-0221/3/08/S08004}}.

\bibitem{Khachatryan:2016bia}
\hrefCMSnoop {}{{CMS Collaboration}, ``{The CMS trigger system}'',} \textit{
  JINST} \textbf{ 12} (2017) P01020,
  \href{http://dx.doi.org/10.1088/1748-0221/12/01/P01020}{\doi{10.1088/1748-0221/12/01/P01020}},
\href{http://www.arXiv.org/abs/1609.02366}{\texttt{arXiv:1609.02366}}.

\bibitem{phase1trackerTDR}
\hrefCMSnoop {}{{CMS Collaboration}, ``{CMS} technical design report for the
  pixel detector upgrade'',} Technical Report CERN-LHCC-2012-016, CMS-TDR-011,
  2012.
\newblock \href{http://dx.doi.org/10.2172/1151650}{\doi{10.2172/1151650}}.

\bibitem{Alwall:2014hca}
J.~Alwall\hrefCMSnoop {}{ {et~al.}, ``The automated computation of tree-level
  and next-to-leading order differential cross sections, and their matching to
  parton shower simulations'',} \textit{ JHEP} \textbf{ 07} (2014) 079,
  \href{http://dx.doi.org/10.1007/JHEP07(2014)079}{\doi{10.1007/JHEP07(2014)079}},
\href{http://www.arXiv.org/abs/1405.0301}{\texttt{arXiv:1405.0301}}.

\bibitem{Nason:2004rx}
\hrefCMSnoop {}{P.~Nason, ``{A new method for combining NLO QCD with shower
  Monte Carlo algorithms}'',} \textit{ JHEP} \textbf{ 11} (2004) 040,
  \href{http://dx.doi.org/10.1088/1126-6708/2004/11/040}{\doi{10.1088/1126-6708/2004/11/040}},
\href{http://www.arXiv.org/abs/hep-ph/0409146}{\texttt{arXiv:hep-ph/0409146}}.

\bibitem{Frixione:2007vw}
\hrefCMSnoop {}{S.~Frixione, P.~Nason, and C.~Oleari, ``{Matching NLO QCD
  computations with parton shower simulations: the \POWHEG method}'',} \textit{
  JHEP} \textbf{ 11} (2007) 070,
  \href{http://dx.doi.org/10.1088/1126-6708/2007/11/070}{\doi{10.1088/1126-6708/2007/11/070}},
\href{http://www.arXiv.org/abs/0709.2092}{\texttt{arXiv:0709.2092}}.

\bibitem{Alioli:2010xd}
\hrefCMSnoop {}{S.~Alioli, P.~Nason, C.~Oleari, and E.~Re, ``{A general
  framework for implementing NLO calculations in shower Monte Carlo programs:
  the \POWHEG BOX}'',} \textit{ JHEP} \textbf{ 06} (2010) 043,
  \href{http://dx.doi.org/10.1007/JHEP06(2010)043}{\doi{10.1007/JHEP06(2010)043}},
\href{http://www.arXiv.org/abs/1002.2581}{\texttt{arXiv:1002.2581}}.

\bibitem{Re:2010bp}
\hrefCMSnoop {}{E.~Re, ``{Single-top $\PW\PQt$-channel production matched with
  parton showers using the POWHEG method}'',} \textit{ Eur. Phys. J. C}
  \textbf{ 71} (2011) 1547,
  \href{http://dx.doi.org/10.1140/epjc/s10052-011-1547-z}{\doi{10.1140/epjc/s10052-011-1547-z}},
\href{http://www.arXiv.org/abs/1009.2450}{\texttt{arXiv:1009.2450}}.

\bibitem{Ball:2011uy}
\hrefCMSnoop {}{{NNPDF} Collaboration, ``{Unbiased global determination of
  parton distributions and their uncertainties at NNLO and at LO}'',} \textit{
  Nucl. Phys. B} \textbf{ 855} (2012) 153,
  \href{http://dx.doi.org/10.1016/j.nuclphysb.2011.09.024}{\doi{10.1016/j.nuclphysb.2011.09.024}},
\href{http://www.arXiv.org/abs/1107.2652}{\texttt{arXiv:1107.2652}}.

\bibitem{Ball:2014uwa}
\hrefCMSnoop {}{{NNPDF} Collaboration, ``{Parton distributions for the LHC Run
  II}'',} \textit{ JHEP} \textbf{ 04} (2015) 040,
  \href{http://dx.doi.org/10.1007/JHEP04(2015)040}{\doi{10.1007/JHEP04(2015)040}},
\href{http://www.arXiv.org/abs/1410.8849}{\texttt{arXiv:1410.8849}}.

\bibitem{Ball:2017nwa}
\hrefCMSnoop {}{{NNPDF} Collaboration, ``{Parton distributions from
  high-precision collider data}'',} \textit{ Eur. Phys. J. C} \textbf{ 77}
  (2017) 663,
  \href{http://dx.doi.org/10.1140/epjc/s10052-017-5199-5}{\doi{10.1140/epjc/s10052-017-5199-5}},
\href{http://www.arXiv.org/abs/1706.00428}{\texttt{arXiv:1706.00428}}.

\bibitem{Sjostrand:2014zea}
T.~Sj{\"o}strand\hrefCMSnoop {}{ {et~al.}, ``An introduction to {PYTHIA}
  8.2'',} \textit{ Comput. Phys. Commun.} \textbf{ 191} (2015) 159,
  \href{http://dx.doi.org/10.1016/j.cpc.2015.01.024}{\doi{10.1016/j.cpc.2015.01.024}},
\href{http://www.arXiv.org/abs/1410.3012}{\texttt{arXiv:1410.3012}}.

\bibitem{Alwall:2007fs}
J.~Alwall\hrefCMSnoop {}{ {et~al.}, ``Comparative study of various algorithms
  for the merging of parton showers and matrix elements in hadronic
  collisions'',} \textit{ Eur. Phys. J. C} \textbf{ 53} (2008) 473,
  \href{http://dx.doi.org/10.1140/epjc/s10052-007-0490-5}{\doi{10.1140/epjc/s10052-007-0490-5}},
\href{http://www.arXiv.org/abs/0706.2569}{\texttt{arXiv:0706.2569}}.

\bibitem{Frederix:2012ps}
\hrefCMSnoop {}{R.~Frederix and S.~Frixione, ``{Merging meets matching in
  MC@NLO}'',} \textit{ JHEP} \textbf{ 12} (2012) 061,
  \href{http://dx.doi.org/10.1007/JHEP12(2012)061}{\doi{10.1007/JHEP12(2012)061}},
\href{http://www.arXiv.org/abs/1209.6215}{\texttt{arXiv:1209.6215}}.

\bibitem{Khachatryan:2015pea}
\hrefCMSnoop {}{{CMS Collaboration}, ``{Event generator tunes obtained from
  underlying event and multiparton scattering measurements}'',} \textit{ Eur.
  Phys. J. C} \textbf{ 76} (2016) 155,
  \href{http://dx.doi.org/10.1140/epjc/s10052-016-3988-x}{\doi{10.1140/epjc/s10052-016-3988-x}},
\href{http://www.arXiv.org/abs/1512.00815}{\texttt{arXiv:1512.00815}}.

\bibitem{Sirunyan:2019dfx}
\hrefCMSnoop {}{{CMS Collaboration}, ``{Extraction and validation of a new set
  of CMS PYTHIA8 tunes from underlying-event measurements}'',} (2019).
  \href{http://www.arXiv.org/abs/1903.12179}{\texttt{arXiv:1903.12179}}.
Submitted to \textit{Eur. Phys. J. C}.

\bibitem{Agostinelli2003250}
\hrefCMSnoop {}{{\GEANTfour} Collaboration, ``{\GEANTfour}~---~a simulation
  toolkit'',} \textit{ Nucl. Instrum. Meth. A} \textbf{ 506} (2003) 250,
\href{http://dx.doi.org/10.1016/S0168-9002(03)01368-8}{\doi{10.1016/S0168-9002(03)01368-8}}.

\bibitem{Abdullin:2011zz}
S.~Abdullin\hrefCMSnoop {}{ {et~al.}, ``{The fast simulation of the CMS
  detector at LHC}'',} \textit{ J. Phys. Conf. Ser.} \textbf{ 331} (2011)
  032049,
\href{http://dx.doi.org/10.1088/1742-6596/331/3/032049}{\doi{10.1088/1742-6596/331/3/032049}}.

\bibitem{Giammanco:2014bza}
\hrefCMSnoop {}{A.~Giammanco, ``{The Fast Simulation of the CMS Experiment}'',}
  \textit{ J. Phys. Conf. Ser.} \textbf{ 513} (2014) 022012,
\href{http://dx.doi.org/10.1088/1742-6596/513/2/022012}{\doi{10.1088/1742-6596/513/2/022012}}.

\bibitem{Li:2012wna}
\hrefCMSnoop {}{Y.~Li and F.~Petriello, ``{Combining QCD and electroweak
  corrections to dilepton production in FEWZ}'',} \textit{ Phys. Rev. D}
  \textbf{ 86} (2012) 094034,
  \href{http://dx.doi.org/10.1103/PhysRevD.86.094034}{\doi{10.1103/PhysRevD.86.094034}},
\href{http://www.arXiv.org/abs/1208.5967}{\texttt{arXiv:1208.5967}}.

\bibitem{Aliev:2010zk}
M.~Aliev\hrefCMSnoop {}{ {et~al.}, ``{HATHOR: HAdronic Top and Heavy quarks
  crOss section calculatoR}'',} \textit{ Comput. Phys. Commun.} \textbf{ 182}
  (2011) 1034,
  \href{http://dx.doi.org/10.1016/j.cpc.2010.12.040}{\doi{10.1016/j.cpc.2010.12.040}},
\href{http://www.arXiv.org/abs/1007.1327}{\texttt{arXiv:1007.1327}}.

\bibitem{Kant:2014oha}
P.~Kant\hrefCMSnoop {}{ {et~al.}, ``{HatHor for single top-quark production:
  Updated predictions and uncertainty estimates for single top-quark production
  in hadronic collisions}'',} \textit{ Comput. Phys. Commun.} \textbf{ 191}
  (2015) 74,
  \href{http://dx.doi.org/10.1016/j.cpc.2015.02.001}{\doi{10.1016/j.cpc.2015.02.001}},
\href{http://www.arXiv.org/abs/1406.4403}{\texttt{arXiv:1406.4403}}.

\bibitem{Beneke:2011mq}
\hrefCMSnoop {}{M.~Beneke, P.~Falgari, S.~Klein, and C.~Schwinn, ``{Hadronic
  top-quark pair production with NNLL threshold resummation}'',} \textit{ Nucl.
  Phys. B} \textbf{ 855} (2012) 695,
  \href{http://dx.doi.org/10.1016/j.nuclphysb.2011.10.021}{\doi{10.1016/j.nuclphysb.2011.10.021}},
\href{http://www.arXiv.org/abs/1109.1536}{\texttt{arXiv:1109.1536}}.

\bibitem{Cacciari:2011hy}
M.~Cacciari\hrefCMSnoop {}{ {et~al.}, ``{Top-pair production at hadron
  colliders with next-to-next-to-leading logarithmic soft-gluon
  resummation}'',} \textit{ Phys. Lett. B} \textbf{ 710} (2012) 612,
  \href{http://dx.doi.org/10.1016/j.physletb.2012.03.013}{\doi{10.1016/j.physletb.2012.03.013}},
\href{http://www.arXiv.org/abs/1111.5869}{\texttt{arXiv:1111.5869}}.

\bibitem{Czakon:2011xx}
\hrefCMSnoop {}{M.~Czakon and A.~Mitov, ``{Top++: A Program for the Calculation
  of the Top-Pair Cross-Section at Hadron Colliders}'',} \textit{ Comput. Phys.
  Commun.} \textbf{ 185} (2014) 2930,
  \href{http://dx.doi.org/10.1016/j.cpc.2014.06.021}{\doi{10.1016/j.cpc.2014.06.021}},
\href{http://www.arXiv.org/abs/1112.5675}{\texttt{arXiv:1112.5675}}.

\bibitem{Baernreuther:2012ws}
\hrefCMSnoop {}{P.~B{\"{a}}rnreuther, M.~Czakon, and A.~Mitov, ``{Percent level
  precision physics at the tevatron: First genuine NNLO QCD corrections to $q
  \bar{q} \to t \bar{t} + X$}'',} \textit{ Phys. Rev. Lett.} \textbf{ 109}
  (2012) 132001,
  \href{http://dx.doi.org/10.1103/PhysRevLett.109.132001}{\doi{10.1103/PhysRevLett.109.132001}},
\href{http://www.arXiv.org/abs/1204.5201}{\texttt{arXiv:1204.5201}}.

\bibitem{Czakon:2012zr}
\hrefCMSnoop {}{M.~Czakon and A.~Mitov, ``{NNLO corrections to top-pair
  production at hadron colliders: the all-fermionic scattering channels}'',}
  \textit{ JHEP} \textbf{ 12} (2012) 054,
  \href{http://dx.doi.org/10.1007/JHEP12(2012)054}{\doi{10.1007/JHEP12(2012)054}},
\href{http://www.arXiv.org/abs/1207.0236}{\texttt{arXiv:1207.0236}}.

\bibitem{Czakon:2012pz}
\hrefCMSnoop {}{M.~Czakon and A.~Mitov, ``{NNLO corrections to top pair
  production at hadron colliders: the quark-gluon reaction}'',} \textit{ JHEP}
  \textbf{ 01} (2013) 080,
  \href{http://dx.doi.org/10.1007/JHEP01(2013)080}{\doi{10.1007/JHEP01(2013)080}},
\href{http://www.arXiv.org/abs/1210.6832}{\texttt{arXiv:1210.6832}}.

\bibitem{Czakon:2013goa}
\hrefCMSnoop {}{M.~Czakon, P.~Fiedler, and A.~Mitov, ``{Total top-quark
  pair-production cross section at hadron colliders through $O(\alpS^4)$}'',}
  \textit{ Phys. Rev. Lett.} \textbf{ 110} (2013) 252004,
  \href{http://dx.doi.org/10.1103/PhysRevLett.110.252004}{\doi{10.1103/PhysRevLett.110.252004}},
\href{http://www.arXiv.org/abs/1303.6254}{\texttt{arXiv:1303.6254}}.

\bibitem{Beenakker:1996ch}
\hrefCMSnoop {}{W.~Beenakker, R.~Hopker, M.~Spira, and P.~M. Zerwas, ``{Squark
  and gluino production at hadron colliders}'',} \textit{ Nucl. Phys. B}
  \textbf{ 492} (1997) 51,
  \href{http://dx.doi.org/10.1016/S0550-3213(97)80027-2}{\doi{10.1016/S0550-3213(97)80027-2}},
\href{http://www.arXiv.org/abs/hep-ph/9610490}{\texttt{arXiv:hep-ph/9610490}}.

\bibitem{Kulesza:2008jb}
\hrefCMSnoop {}{A.~Kulesza and L.~Motyka, ``{Threshold resummation for
  squark-antisquark and gluino-pair production at the LHC}'',} \textit{ Phys.
  Rev. Lett.} \textbf{ 102} (2009) 111802,
  \href{http://dx.doi.org/10.1103/PhysRevLett.102.111802}{\doi{10.1103/PhysRevLett.102.111802}},
\href{http://www.arXiv.org/abs/0807.2405}{\texttt{arXiv:0807.2405}}.

\bibitem{Kulesza:2009kq}
\hrefCMSnoop {}{A.~Kulesza and L.~Motyka, ``{Soft gluon resummation for the
  production of gluino-gluino and squark-antisquark pairs at the LHC}'',}
  \textit{ Phys. Rev. D} \textbf{ 80} (2009) 095004,
  \href{http://dx.doi.org/10.1103/PhysRevD.80.095004}{\doi{10.1103/PhysRevD.80.095004}},
\href{http://www.arXiv.org/abs/0905.4749}{\texttt{arXiv:0905.4749}}.

\bibitem{Beenakker:2009ha}
W.~Beenakker\hrefCMSnoop {}{ {et~al.}, ``{Soft-gluon resummation for squark and
  gluino hadroproduction}'',} \textit{ JHEP} \textbf{ 12} (2009) 041,
  \href{http://dx.doi.org/10.1088/1126-6708/2009/12/041}{\doi{10.1088/1126-6708/2009/12/041}},
\href{http://www.arXiv.org/abs/0909.4418}{\texttt{arXiv:0909.4418}}.

\bibitem{Beenakker:2011fu}
W.~Beenakker\hrefCMSnoop {}{ {et~al.}, ``{Squark and Gluino
  Hadroproduction}'',} \textit{ Int. J. Mod. Phys. A} \textbf{ 26} (2011) 2637,
  \href{http://dx.doi.org/10.1142/S0217751X11053560}{\doi{10.1142/S0217751X11053560}},
\href{http://www.arXiv.org/abs/1105.1110}{\texttt{arXiv:1105.1110}}.

\bibitem{Borschensky:2014cia}
C.~Borschensky\hrefCMSnoop {}{ {et~al.}, ``{Squark and gluino production cross
  sections in $pp$ collisions at $\sqrt{s} =$13, 14, 33 and $100\TeV$}'',}
  \textit{ Eur. Phys. J. C} \textbf{ 74} (2014) 3174,
  \href{http://dx.doi.org/10.1140/epjc/s10052-014-3174-y}{\doi{10.1140/epjc/s10052-014-3174-y}},
\href{http://www.arXiv.org/abs/1407.5066}{\texttt{arXiv:1407.5066}}.

\bibitem{Beenakker:2016lwe}
W.~Beenakker\hrefCMSnoop {}{ {et~al.}, ``{NNLL-fast: predictions for coloured
  supersymmetric particle production at the LHC with threshold and Coulomb
  resummation}'',} \textit{ JHEP} \textbf{ 12} (2016) 133,
  \href{http://dx.doi.org/10.1007/JHEP12(2016)133}{\doi{10.1007/JHEP12(2016)133}},
\href{http://www.arXiv.org/abs/1607.07741}{\texttt{arXiv:1607.07741}}.

\bibitem{Sirunyan:2017ulk}
\hrefCMSnoop {}{{CMS Collaboration}, ``Particle-flow reconstruction and global
  event description with the {CMS} detector'',} \textit{ JINST} \textbf{ 12}
  (2017) P10003,
  \href{http://dx.doi.org/10.1088/1748-0221/12/10/P10003}{\doi{10.1088/1748-0221/12/10/P10003}},
\href{http://www.arXiv.org/abs/1706.04965}{\texttt{arXiv:1706.04965}}.

\bibitem{Cacciari:2005hq}
\hrefCMSnoop {}{M.~Cacciari and G.~P. Salam, ``Dispelling the {$N^{3}$} myth
  for the {$\kt$} jet-finder'',} \textit{ Phys. Lett. B} \textbf{ 641} (2006)
  57,
  \href{http://dx.doi.org/10.1016/j.physletb.2006.08.037}{\doi{10.1016/j.physletb.2006.08.037}},
\href{http://www.arXiv.org/abs/hep-ph/0512210}{\texttt{arXiv:hep-ph/0512210}}.

\bibitem{Cacciari:2008gp}
\hrefCMSnoop {}{M.~Cacciari, G.~P. Salam, and G.~Soyez, ``The anti-{$\kt$} jet
  clustering algorithm'',} \textit{ JHEP} \textbf{ 04} (2008) 063,
  \href{http://dx.doi.org/10.1088/1126-6708/2008/04/063}{\doi{10.1088/1126-6708/2008/04/063}},
  \href{http://www.arXiv.org/abs/0802.1189}{\texttt{arXiv:0802.1189}}.

\bibitem{Cacciari:2011ma}
\hrefCMSnoop {}{M.~Cacciari, G.~P. Salam, and G.~Soyez, ``{FastJet} user
  manual'',} \textit{ Eur. Phys. J. C} \textbf{ 72} (2012) 1896,
  \href{http://dx.doi.org/10.1140/epjc/s10052-012-1896-2}{\doi{10.1140/epjc/s10052-012-1896-2}},
\href{http://www.arXiv.org/abs/1111.6097}{\texttt{arXiv:1111.6097}}.

\bibitem{Chatrchyan:2011tn}
\hrefCMSnoop {}{{CMS Collaboration}, ``{Missing transverse energy performance
  of the CMS detector}'',} \textit{ JINST} \textbf{ 6} (2011) P09001,
  \href{http://dx.doi.org/10.1088/1748-0221/6/09/P09001}{\doi{10.1088/1748-0221/6/09/P09001}},
\href{http://www.arXiv.org/abs/1106.5048}{\texttt{arXiv:1106.5048}}.

\bibitem{Khachatryan:2015hwa}
\hrefCMSnoop {}{{CMS Collaboration}, ``{Performance of electron reconstruction
  and selection with the CMS detector in proton-proton collisions at
  $\sqrt{s}=8\TeV$}'',} \textit{ JINST} \textbf{ 10} (2015) P06005,
  \href{http://dx.doi.org/10.1088/1748-0221/10/06/P06005}{\doi{10.1088/1748-0221/10/06/P06005}},
\href{http://www.arXiv.org/abs/1502.02701}{\texttt{arXiv:1502.02701}}.

\bibitem{Sirunyan:2018fpa}
\hrefCMSnoop {}{{CMS Collaboration}, ``Performance of the {CMS} muon detector
  and muon reconstruction with proton-proton collisions at
  {$\sqrt{s}=13\TeV$}'',} \textit{ JINST} \textbf{ 13} (2018) P06015,
  \href{http://dx.doi.org/10.1088/1748-0221/13/06/P06015}{\doi{10.1088/1748-0221/13/06/P06015}},
\href{http://www.arXiv.org/abs/1804.04528}{\texttt{arXiv:1804.04528}}.

\bibitem{cacciari-2008-659}
\hrefCMSnoop {}{M.~Cacciari and G.~P. Salam, ``Pileup subtraction using jet
  areas'',} \textit{ Phys. Lett. B} \textbf{ 659} (2008) 119,
  \href{http://dx.doi.org/10.1016/j.physletb.2007.09.077}{\doi{10.1016/j.physletb.2007.09.077}},
\href{http://www.arXiv.org/abs/0707.1378}{\texttt{arXiv:0707.1378}}.

\bibitem{jetid13TeV}
\href {https://cds.cern.ch/record/2256875}{{CMS Collaboration}, ``{Jet
  algorithms performance in 13 TeV data}'',} CMS Physics Analysis Summary
  CMS-PAS-JME-16-003, 2017.

\bibitem{Khachatryan:2016kdb}
\hrefCMSnoop {}{{CMS Collaboration}, ``{Jet energy scale and resolution in the
  CMS experiment in $\Pp\Pp$ collisions at 8 TeV}'',} \textit{ JINST} \textbf{
  12} (2017) P02014,
  \href{http://dx.doi.org/10.1088/1748-0221/12/02/P02014}{\doi{10.1088/1748-0221/12/02/P02014}},
\href{http://www.arXiv.org/abs/1607.03663}{\texttt{arXiv:1607.03663}}.

\bibitem{CMS-DP-2018-028}
\href {http://cds.cern.ch/record/2622157}{{CMS Collaboration}, ``{Jet energy
  scale and resolution performance with 13 TeV data collected by CMS in
  2016}'',} {Detector Performance Report} CMS-DP-2018-028, 2018.

\bibitem{Sirunyan:2017ezt}
\hrefCMSnoop {}{{CMS Collaboration}, ``{Identification of heavy-flavour jets
  with the CMS detector in $\Pp\Pp$ collisions at 13 TeV}'',} \textit{ JINST}
  \textbf{ 13} (2018) P05011,
  \href{http://dx.doi.org/10.1088/1748-0221/13/05/P05011}{\doi{10.1088/1748-0221/13/05/P05011}},
\href{http://www.arXiv.org/abs/1712.07158}{\texttt{arXiv:1712.07158}}.

\bibitem{Khachatryan:2011wq}
\hrefCMSnoop {}{{CMS Collaboration}, ``{Measurement of $B\bar{B}$ Angular
  Correlations based on Secondary Vertex Reconstruction at $\sqrt{s}=7$
  TeV}'',} \textit{ JHEP} \textbf{ 03} (2011) 136,
  \href{http://dx.doi.org/10.1007/JHEP03(2011)136}{\doi{10.1007/JHEP03(2011)136}},
\href{http://www.arXiv.org/abs/1102.3194}{\texttt{arXiv:1102.3194}}.

\bibitem{ganin2014unsupervised}
\href {https://arxiv.org/pdf/1409.7495v2.pdf}{Y.~Ganin and V.~Lempitsky,
  ``Unsupervised domain adaptation by backpropagation'',} 2014.
\newblock \url {https://arxiv.org/pdf/1409.7495v2.pdf}.

\bibitem{CMS-PAS-JME-18-002}
\href {http://cds.cern.ch/record/2683870}{{CMS Collaboration}, ``{Machine
  learning-based identification of highly Lorentz-boosted hadronically decaying
  particles at the CMS experiment}'',} {CMS Physics Analysis Summary}
  CMS-PAS-JME-18-002, 2019.

\bibitem{Graesser:2012qy}
\hrefCMSnoop {}{M.~L. Graesser and J.~Shelton, ``Hunting mixed top squark
  decays'',} \textit{ Phys. Rev. Lett.} \textbf{ 111} (2013) 121802,
  \href{http://dx.doi.org/10.1103/PhysRevLett.111.121802}{\doi{10.1103/PhysRevLett.111.121802}},
\href{http://www.arXiv.org/abs/1212.4495}{\texttt{arXiv:1212.4495}}.

\bibitem{Sirunyan:2017uzs}
\hrefCMSnoop {}{{CMS Collaboration}, ``{Measurement of the cross section for
  top quark pair production in association with a $\PW$ or $\PZ$ boson in
  proton-proton collisions at $\sqrt{s}=13\TeV$}'',} \textit{ JHEP} \textbf{
  08} (2018) 011,
  \href{http://dx.doi.org/10.1007/JHEP08(2018)011}{\doi{10.1007/JHEP08(2018)011}},
\href{http://www.arXiv.org/abs/1711.02547}{\texttt{arXiv:1711.02547}}.

\bibitem{Catani:2003zt}
\hrefCMSnoop {}{S.~Catani, D.~de~Florian, M.~Grazzini, and P.~Nason,
  ``{Soft-gluon resummation for Higgs boson production at hadron colliders}'',}
  \textit{ JHEP} \textbf{ 07} (2003) 028,
  \href{http://dx.doi.org/10.1088/1126-6708/2003/07/028}{\doi{10.1088/1126-6708/2003/07/028}},
\href{http://www.arXiv.org/abs/hep-ph/0306211}{\texttt{arXiv:hep-ph/0306211}}.

\bibitem{Cacciari2003fi}
M.~Cacciari\hrefCMSnoop {}{ {et~al.}, ``{The \ttbar cross-section at $1.8\TeV$
  and $1.96\TeV$: a study of the systematics due to parton densities and scale
  dependence}'',} \textit{ JHEP} \textbf{ 04} (2004) 068,
  \href{http://dx.doi.org/10.1088/1126-6708/2004/04/068}{\doi{10.1088/1126-6708/2004/04/068}},
\href{http://www.arXiv.org/abs/hep-ph/0303085}{\texttt{arXiv:hep-ph/0303085}}.

\bibitem{Butterworth:2015oua}
\hrefCMSnoop {}{J.~Butterworth {et~al.}, ``{PDF4LHC recommendations for LHC Run
  II}'',} \textit{ J. Phys. G} \textbf{ 43} (2016) 023001,
  \href{http://dx.doi.org/10.1088/0954-3899/43/2/023001}{\doi{10.1088/0954-3899/43/2/023001}},
\href{http://www.arXiv.org/abs/1510.03865}{\texttt{arXiv:1510.03865}}.

\bibitem{Junk:1999kv}
\hrefCMSnoop {}{T.~Junk, ``{Confidence level computation for combining searches
  with small statistics}'',} \textit{ Nucl. Instrum. Meth. A} \textbf{ 434}
  (1999) 435,
  \href{http://dx.doi.org/10.1016/S0168-9002(99)00498-2}{\doi{10.1016/S0168-9002(99)00498-2}},
\href{http://www.arXiv.org/abs/hep-ex/9902006}{\texttt{arXiv:hep-ex/9902006}}.

\bibitem{Read:2002hq}
\hrefCMSnoop {}{A.~L. Read, ``Presentation of search results: the {$CL_{S}$}
  technique'',} \textit{ J. Phys. G} \textbf{ 28} (2002) 2693,
  \href{http://dx.doi.org/10.1088/0954-3899/28/10/313}{\doi{10.1088/0954-3899/28/10/313}}.

\bibitem{Cowan:2010js}
\hrefCMSnoop {}{G.~Cowan, K.~Cranmer, E.~Gross, and O.~Vitells, ``Asymptotic
  formulae for likelihood-based tests of new physics'',} \textit{ Eur. Phys. J.
  C} \textbf{ 71} (2011) 1554,
  \href{http://dx.doi.org/10.1140/epjc/s10052-011-1554-0}{\doi{10.1140/epjc/s10052-011-1554-0}},
  \href{http://www.arXiv.org/abs/1007.1727}{\texttt{arXiv:1007.1727}}.
  [Erratum: 10.1140/epjc/s10052-013-2501-z].

\bibitem{ATLAS:2011tau}
\href {http://cds.cern.ch/record/1379837}{{ATLAS and CMS Collaborations, and
  the LHC Higgs Combination Group}, ``Procedure for the {LHC} {Higgs} boson
  search combination in summer 2011'',} ATLAS/CMS joint note
  ATL-PHYS-PUB-2011-011, CMS-NOTE-2011-005, 2011.

\end{thebibliography}\endgroup
\cleardoublepage \appendix\section{The CMS Collaboration \label{app:collab}}\begin{sloppypar}\hyphenpenalty=5000\widowpenalty=500\clubpenalty=5000\vskip\cmsinstskip
\textbf{Yerevan Physics Institute, Yerevan, Armenia}\\*[0pt]
A.M.~Sirunyan$^{\textrm{\dag}}$, A.~Tumasyan
\vskip\cmsinstskip
\textbf{Institut f\"{u}r Hochenergiephysik, Wien, Austria}\\*[0pt]
W.~Adam, F.~Ambrogi, T.~Bergauer, J.~Brandstetter, M.~Dragicevic, J.~Er\"{o}, A.~Escalante~Del~Valle, M.~Flechl, R.~Fr\"{u}hwirth\cmsAuthorMark{1}, M.~Jeitler\cmsAuthorMark{1}, N.~Krammer, I.~Kr\"{a}tschmer, D.~Liko, T.~Madlener, I.~Mikulec, N.~Rad, J.~Schieck\cmsAuthorMark{1}, R.~Sch\"{o}fbeck, M.~Spanring, D.~Spitzbart, W.~Waltenberger, C.-E.~Wulz\cmsAuthorMark{1}, M.~Zarucki
% ...
\vskip\cmsinstskip
\dag: Deceased\\
1:  Also at Vienna University of Technology, Vienna, Austria\\
% ...
\end{sloppypar}
\end{document}